# The Group Structure Bases of a Foundational Approach to Physics

## Peter Rowlands[1], J. P. Cullerne[2], and Brian D. Koberlein[3]

[1] *IQ Group and Science Communication Unit, Department of Physics, University of Liverpool, Oliver Lodge Laboratory, Oxford Street, P.O. Box 147, Liverpool, L69 7ZE, UK. e-mail prowl@hep.ph.liv.ac.uk and prowl@csc.liv.uk*

[2] *IQ Group, Department of Computer Science, University of Liverpool, Chadwick Laboratory, Peach Street, Liverpool, L69 7ZF, UK. e-mail jpc@wincoll.ac.uk*

[3] *Department of Physics, SUNY Geneseo State College, 1 College Circle, Geneseo, NY 14454, USA. e-mail Kober@geneseo.edu and Brian.Koberlein@esc.edu*

*Abstract* Based on a fundamental symmetry between space, time, mass and charge, a series of group structures of physical interest is generated, ranging from $C_2$ to $E_8$. The most significant result of this analysis is a version of the Dirac equation combining quaternions and multivariate 4-vectors, which is already second quantized and intrinsically supersymmetric, and which automatically leads to a symmetry breaking, with the creation of specific particle structures and a mass-generating mechanism.

## 1 Introduction

Physics appears to be an attempt at deriving something from nothing. There is no evidence to suppose that what we describe as 'reality', either in the concrete sense of something directly perceived, or in the sense of an ontological concept beyond immediate perception, has any defining characteristic. In fact it would appear that 'reality' or 'nature' goes out of its way to avoid being characterized. Physicists have learned to deal with this by default, effectively by a natural selection of the only method which works. This is to insert a probe into nature, observe how the response denies validity to the probe, and then incorporate both probe and response under the guise of 'symmetry'.[1]

Inserting a probe is the process we define as 'measurement' or observation. Measurability depends on discreteness. The simplest discrete thing is a point. So we start with the creation of a point or point 'particle'. A single particle has to have inherent symmetries. The most fundamental belong to the four parameters space, time, mass and charge (if we take charge in its most general sense as the source of strong and weak, as well as electromagnetic interactions). Space and time are the only true ways of approaching the concept of variation in nature, while mass and charge provide the conserved sources of the four known interactions: gravity; electromagnetism; and the strong and weak nuclear forces. It has never been demonstrated that anything in physics is more fundamental than these concepts, and it has never been demonstrated that we can construct physics on any other basis. We will now show that these fundamental concepts are not independently defined but have fundamental symmetries which connect them, linked with three distinct properties. Parameters may be conserved or nonconserved; orderable or nonorderable; countable or noncountable. Each parameter takes on one of each of these properties or its symmetric (and absolute) opposite, and the four parameters, taken together on the basis of these properties, form a group. The symmetry constrains each parameter to have *one* property in common with each other, and two which are entirely opposite.[2-6] It will be necessary to our argument to give a detailed discussion of the different aspects of this symmetry.

## 2 Conserved or nonconserved

The conservation laws are the most fundamental in physics, and they are the source of its universality. The most fundamental principles of physics are essentially equivalent to statements that some quantities are absolutely conserved while others are allowed to vary without restriction; and all such conservation principles are derived ultimately from the conservation properties of mass and charge. Fundamental physics assumes that the conservation of electric charge is true in all conceivable circumstances, and such properties of fundamental particles as lepton and baryon conservation suggests that the same principle applies also to the sources of the strong and weak interactions. Mass (in the sense of mass-energy, or the source of the gravitational force) is also absolutely conserved; and all these conservation laws are *local*, applying to the amount of each quantity at a given place in a given time. Classically, we give each element of mass or charge an *identity* which it retains throughout all interactions, subject only, in the case of charge, to its annihilation by an element with the opposite sign; and, although the identity of individual particles is not maintained in quantum mechanics, we still retain the requirement of local conservation.

Since mass and charge are alike in their absolute conservation, space and time are constrained to be opposite in their absolute *nonconservation*. This manifests itself in a number of ways. The *nonidentity* of the elements of space and time is manifested in their property of *translation symmetry*. So each element of space and time is exactly like every other, and physical equations must be structured to incorporate this. Two very significant physical principles follow from this via Noether's theorem, the mathematical result which states that, for every global transformation preserving the Lagrangian density, there exists a conserved quantity. Thus, the translation symmetry of time requires, and is identical with, the conservation of energy, while the translation symmetry of space requires, and is identical with, the conservation of linear momentum. Space, as a three-dimensional quantity, also has rotation symmetry, or nonidentity of spatial *direction*, which Noether's theorem equates to the conservation of angular momentum.

By analogy, to show the exact oppositeness of the conserved quantities, we may apply a concept of translation *a*symmetry to both mass and charge, as another expression of local conservation. Each element of mass and charge is untranslatable to any other. Such conserved quantities can, of course, only be defined with respect to changes in the nonconserved quantities, and what we define as interactions can often be thought of as a statement of all those possible changes in the nonconserved quantities which will maintain the value of the conserved ones. We look at what remains invariant (conserved) under certain groups of transformations (nonconserved). The aspects of identical particle wavefunctions which are truly interchangeable in quantum mechanics are the nonconserved space and time coordinates rather than the conserved charges.

The conserved quantities, of course, include composite ones such as energy, momentum and angular momentum, as well as the more elementary ones, mass and charge, but the conservation of the composite ones depends directly on the conservation of the more elementary ones, and each of the composite conservation laws relates also to one of the aspects of the *nonconservation* of space or time. The behaviour of *all* physical systems is then described entirely in terms of a combination of conservation and nonconservation principles, the nonconservation of space and time usually requiring them to be expressed by differentials.

Gauge invariance, as used in both classical and quantum physics, is the classic expression of how a system remains conservative under arbitrary changes in the coordinates which do not produce changes in the values of conserved quantities such as charge, energy, momentum and angular momentum; and quantum mechanics is simply the ultimate expression of the arbitrariness of spatial and temporal coordinates, which are subject to absolute and arbitrary change. Significantly, the Yang-Mills principle used in particle physics requires nonconservation, in the form of arbitrary phase changes, to be local in exactly the same way as conservation.

## 3 Orderable or nonorderable

Although Minkowski, in 1908, in mathematically uniting space and time coordinates in what we now call a 4-vector, proclaimed that: 'From now on, space by itself, and time by itself, are destined to sink into shadows, and only a kind of union of both to retain an independent existence',[7] the very structure of the 4-vector proves that the two concepts retain significant physical differences. Thus, while Pythagorean addition produces positive values for the squares of the three spatial dimensions, the squared value of time becomes negative in the Minkowski 4-vector. In other words, time is represented by an imaginary or nonorderable number. Only its square has an orderable value; the time coordinate itself does not. This imaginary representation is often claimed as merely 'convenient', but there is no obvious reason why it should be convenient to do such a non-physical thing. However, when we also realize that an imaginary representation for time also makes uniform velocity imaginary, while acceleration remains real, we can see that there might be other physical reasons for using it.

The parallel representation of mass and charge is also instructive. Force terms require squared masses and charges, which we can imagine as being combined in the same way as the squared space and time terms in Pythagorean addition. However, there remains the unexplained fact that forces between like masses are attractive, whereas forces between like (electric) charges are repulsive; that is, the forces between like masses and like charges, and hence the squared values of those masses and charges, have opposite signs. If, however, we choose to represent charges by imaginary numbers and masses by real ones, we create a symmetrical representation for the Newton and Coulomb force laws:

$$F = -\frac{Gm_1m_2}{r^2}$$

$$F = -\frac{iq_1iq_2}{4\pi\varepsilon_0 r^2}$$

In addition, the other two charged-based forces, the strong and weak interactions, are like the electric force in being repulsive for like particles, and so the source terms for these forces are also presumably defined by other imaginary numbers. The natural mathematical description for this arrangement is the quaternion system, discovered in 1843, in which $i, j$ and $k$, the three square roots of $-1$, are related by the formulae:

$$i^2 = j^2 = k^2 = ijk = -1 .$$

With mass taking up the role of the real part, the charge-mass quaternion becomes the exact symmetrical opposite of the space-time 4-vector, with three imaginary parts and one real one (an ordinary real number), as opposed to three real parts and one imaginary. The quaternions, which break the rule of commutativity, also have extra significance in being unique. No other extension of ordinary complex algebra to incorporate multiple imaginary dimensions is possible, unless we are also prepared to break associativity, and, even then, the 8-part octonions or Cayley numbers, which break associativity, are the only remaining division algebra after the real and complex numbers and quaternions. In other words, if we wish to have an associative division algebra, which includes nonorderable (imaginary) multidimensional quantities, then we are restricted precisely to a 3-dimensional system. Hamilton, who first discovered the quaternions after finding that a system with two imaginary parts was impossible, felt that he had discovered in them the explanation for the 3-dimensionality space, with time taking up the fourth or real part. According to our analysis, quaternions are more conveniently applied to the three imaginary components of charge, with mass taking up the real part. However, space and time then become a three real- and one imaginary-part system by *symmetry*. The three components of charge (say, $ie, js, kw$) can, in this way, be considered as the 'dimensions' of a single charge parameter, with their squared values used in the calculation of forces added, in the same way as the three parts of space, by Pythagorean addition:

| space-time | $\mathbf{i}x$ | $\mathbf{j}y$ | $\mathbf{k}z$ | $it$ |
|------------|------|------|------|------|
| mass-charge | $\mathbf{i}e$ | $\mathbf{j}s$ | $\mathbf{k}w$ | $m$ |

An important consequence of making this symmetry exact is that the vector property of space must be extended to incorporate a quaternionic-like 'full' product between two vectors, combining the scalar product with $i$ times the vector product. The extra vector terms in this product are responsible for the otherwise 'mysterious' spin property in quantum mechanics.

Though charge, like space, is multidimensional, charge dimensions will show differences to spatial dimensions, since charge is a conserved quantity, unlike space. In particular, we can expect conservation in dimension as well as in quantity. Charge should exhibit a rotation *a*symmetry, in which the sources of the electromagnetic, weak and strong interactions are separately conserved, and incapable of interconversion. (This does not affect the Weinberg-Salam unification of electromagnetic and weak forces, which is a statement of the identity of effect in the two interactions, under ideal conditions, not of identity of the sources; the three quaternion operators $i, j$ and $k$ are different sources, though identical in effect.) Baryon and lepton conservation are obvious consequences, since baryons are the only particles with strong, as well as weak, components, and leptons are the only particles with weak, but no strong, components. The proton, also, which has a strong charge measured by its baryon number, will not be able to decay to products like the positron and neutral pion, which have none.

Various further benefits also result from the nonorderable nature of imaginary numbers. One is that mass, which is real in itself as well as in its squared value, has two physical manifestations: inertia, which is the property of the linear term, and gravitation, which is the property of the squared value. We can detect a single mass through the property of inertia, independent of its interaction with other masses, whereas a charge is only detectable in the presence of another charge. Another benefit is that real numbers, by their property of orderability, allow positive (or negative) solutions to be privileged in algebraic equations; imaginary numbers, as nonorderable quantities, do not. Hence, every complex (or quaternion) equation which has a positive solution also has an algebraically indistinguishable negative solution (the complex or quaternion conjugate). Thus, though we can use the real nature of mass to privilege the positive value in algebraic equations, we cannot choose a positive solution for charge without allowing the equal possibility of the negative one. The consequence of this is the existence of antiparticles, for even those particles, such as the neutron and neutrino, which have no electric charge still have antiparticles because they have strong and / or weak charges whose signs may be reversed.

## 4 Countable or noncountable

It is important that we recognize that only space is ever measured. All observable objects define a space, and so any object can be used, in principle, for measuring space, without setting up any special conditions. This is not true of time. It is often claimed that clocks are devices used for measuring time, but measurement implies repetition and time, as an irreversible parameter, cannot be repeated. In fact all clocks use some concept of the repetition of a *spatial* interval, and require very special conditions and assumptions to make even this possible.

The ancient paradoxes of Zeno of Elea are directly concerned with the differences between space and time. One famous example is the race between Achilles and the Tortoise, where Achilles can never catch up if he gives the Tortoise a start, no matter how much faster he is. Another is the Dichotomy Paradox, in which an object moving over any distance can never get started because it must cover half the distance before it covers the whole, and a quarter of the distance before it covers half, and so on; thus, an infinite number of operations are required for the object to move any distance in a finite amount of time.

The problems arise, as numerous philosophers have realized, from the idea that time must be infinitely divisible, like space. However, it is much more likely that time, unlike space, is an absolute continuum, without measurable instants; that is, it is truly analogue, and the mental category of absolute continuity exists precisely because of this. This is apparently what, for example, Henri Bergson believed, but he also believed, as we do not, that time is 'fundamentally irreducible to mathematical terms'.[8] This seems to be untrue because a mathematics involving absolute continuity would appear to be possible in terms of Cantor's definition of the real numbers. The numbers representing time are noncountable, but they can be represented in mathematical terms.

However, absolute continuity is not compatible with indefinite divisibility, such as we perceive in space, in either mathematical or philosophical terms. The very divisibility of space is what denies it *absolute* continuity; the elastic nature of the divisibility comes from the entirely different property of nonconservation. Space, as a countable or noncontinuous quantity, will certainly have units, but, because it is also a nonconserved quantity, these units will not be fixed; and the whole process of measurement through space depends crucially on these two properties.

The countability of space requires it to be discontinuous in both quantity and direction, to be reversible and capable of changes in orientation; without these properties, practical measurement would be impossible. The irreversibility of time, however, is precisely the same property as its noncountability or absolute continuity. Reversal cannot be accomplished without discontinuity, nor can changes in orientation. So it is clear that time, if it is to be absolutely continuous, cannot have multiple dimensions. The so-called 'reversibility paradox' has time reversible in mathematical sign, according to nearly all the laws of physics, but not reversible in physical consequences. This, however, is a result of using imaginary numbers to characterize time. Imaginary numbers are not privileged according to sign, and so must allow equal validity to positive and negative mathematical solutions, but the property of continuity or noncountability ensures that the physical apprehension of time will be in a single direction.

The noncountability of time is also responsible for the fact that time, in the definition of velocity and acceleration, the basic quantities used in dynamics, is the independent variable, whereas space is the dependent variable. We have no control over the variation of time, as we have of space, and so its variation is necessarily independent.

For symmetry, we require one other parameter to be continuous, or noncountable, like time; this is mass (or mass-energy, the source of gravitational interactions). This is present at every point in space in the form of the filled vacuum or Higgs field, and its signature is the unipolarity of the parameter or the absence of negative mass, just as the signature of time's continuity is its irreversibility. It is also, as we will see, the source of nonlocality in quantum mechanics. The unipolarity of mass is the reason why we have a CPT, rather than an MCPT, theorem, C standing for charge conjugation, P for space reflection and T for time reversal, all of which have two mathematical sign options.

This means that the remaining parameter, charge, must be discontinuous or countable, like space. Charge is, of course, observed in discrete units, which, unlike those of space, are fixed, because charge is also a conserved quantity. It is also, like space, multidimensional. Multidimensionality, as we have seen, cannot be a property of continuous quantities, but there are also direct arguments for suggesting that it must be an essential property of discrete ones. In principle, we cannot demonstrate discreteness in a one-dimensional system. A line, for example, which is often used as an illustration of one-dimensionality, can, in fact, be constructed only within a system which is at least two-dimensional. Truly one-dimensional space would produce only a point with no extension. We couldn't use it to demonstrate discreteness, and certainly not the variable discreteness characteristic of space.

It seems to be multidimensionality as such, rather than any particular degree of multidimensionality which is responsible for creating the additional level of discreteness required by the introduction of algebraic numbers, and even of transcendental numbers such as $\pi$; two independent dimensions are all that we require to create the required level of incommensurability at the rational number level. So it would seem that the introduction of a third dimension requires no qualitatively new type of number. Also, although only the rational numbers are needed to define the measurement process using a single dimension, the *nonconserved* nature of space means that the units can be redefined any way we like and so we may well need some version of constructed reals to link all possible systems of units, and to express this degree of freedom.

Dimensionality, as such, may be defined as the use of Pythagorean addition, or the addition of the squared values of numbers, as a separate process from ordinary addition. In this sense, all the parameters are dimensional, and dimensionality may be considered as an ultimate result of the dualities evident in the fundamental group symmetry, such as the co-existence of symmetrical real and imaginary quantities. The process of squaring becomes related to the process of doubling, and the process of square-rooting becomes related to division by the factor 2, which is itself related to the creation of the mathematical concepts of $-1 + 1 = 0$ and $1 + 1 = 2$. The creation of the original discrete or point-like 'probe' and its symmetrical negation can be considered as a process of generating something from nothing which is necessarily dual in its outcome. We create something, but the totality of the something and its symmetrical opposite reduces it to nothing. In fact, to generate something from nothing, our point-particle probes have to become nilpotents, which are nonreal in themselves, and which square to zero in their only real manifestation.

The dimensionality of charges and masses has an important physical manifestation. Charge and mass elements, unlike those of space and time, are individual and conserved, and so the process of 'squaring' becomes a universal 'interaction' between each individual element and each other, which is intrinsically nonlocal, and which manifests itself in the classical concept of force and in the quantum mechanical wavefunction. At the same time, our use of the in-built 4-vector connection between space and time will ensure that the transfer of energy between discrete particles of matter remains time-delayed by the constant relating space and time units in their mathematical combination.

The physical difference between space and time has significant consequences when we combine them dimensionally in the Minkowski 4-vector. We are obliged to either make time space-like (or discrete) or to make space time-like (or continuous), and so we automatically generate dualistic theories. Classic examples of choosing the discrete option are particle theories, special relativity and Heisenberg's quantum mechanics, while the corresponding results of the continuous options are wave theories, Lorentzian relativity and Schrödinger's wave mechanics.[9-11] In the Heisenberg theory, mass and time effectively become discrete quantities, while Schrödinger makes space and charge continuous in the wavefunction. These options are only true of the mathematical structure, for the process of measurement respectively restores continuous mass via the uncertainty principle and the virtual vacuum, and discrete space and charge via the collapse of the wavefunction.

The fundamental distinction between the physical status of space and time has consequences for mathematics as well as for physics. From the beginning, there were two processes of differentiation: a discrete one, modelled on variation in space, using infinitesimals; and a continuous one, modelled on variation in time, defined by a limit process. Mathematically, neither option is any more valid than the other, for differentiation is a property linked to nonconservation, and not concerned, in principle, with the difference between absolute continuity and indefinite divisibility. However, in physical terms, there are differences, for example in the solutions of Zeno's paradoxes, which require the continuous concept of a limiting process to achieve a valid result.

In more recent times, it has become apparent that the Cantorian definition of an absolutely continuous set of real numbers cannot be applied to space, which is not continuous but infinitely divisible. It would seem that, to describe space, we must use Skolem's 'non-standard' arithmetic of 1934, with its denumerable model of the reals, and the non-Archimedean geometry, which relates this to space, in addition to the non-standard analysis of Abraham Robinson, which treats infinitesimals as though they had the properties of real numbers. These versions of non-standard mathematics are a reflection of the discrete nature of space while 'standard' results (based on limiting processes) rely on the continuity of time. Both methods give valid proofs for purely mathematical theorems, because both are valid in a purely mathematical sense, and the validity of the non-standard methods is established by the Löwenheim-Skolem theorem, which states that any consistent finite, formal theory has a denumerable model, with the elements of its domain in a one-to-one correspondence with the positive integers. It would seem that available mathematical options arise simultaneously with the available physical options, and ultimately stem from physical symmetries. Continuity and discontinuity, finiteness and infinity, and so on, would not exist as mathematical categories unless they were also physical categories.

A further significant aspect of nonstandard analysis is that it has been a key ingredient in developing topos theory, which provides a view of space as arising from some kind of mathematical structure and possibly including dynamism, the points of the space being infinitesimal (but smeared) and nilpotent, or square roots of zero. Topos theory offers very attractive possibilities for formalizing some of the points made here. Space, in the present theory, arises from the process of counting or measurement; dimensionality is a necessary consequence of discreteness (or smearing); dynamism, with time, mass and charge emerging automatically from the symmetry, is in this sense in-built; and, as we will see later, the nilpotent wavefunction which we will construct for the pointlike charges could be related to the nilpotents needed to construct the points in space, especially as the nilpotent wavefunction creates an infinite series of further nilpotents in the continuous vacuum.

## 5 Object group structure

From the previous sections, it would appear that the four basic parameters are distributed between three sets of opposing paired categories: real / imaginary (alternatively, orderable / nonorderable), conserved / nonconserved, countable / noncountable (alternatively, discrete / continuous, divisible / indivisible). Each parameter is paired off with a different partner in each of the categories, according to the following scheme:

| | | | |
|---|---|---|---|
| **space** | real | nonconserved | countable |
| **time** | imaginary | nonconserved | noncountable |
| **mass** | real | conserved | noncountable |
| **charge** | imaginary | conserved | countable |

Where properties match, they are identical, and where they oppose, they are in exact opposition. Each pair forms an abstract group, and therefore defines the symmetry of a physical object. The pairs (orderable / nonorderable) and (conserved / nonconserved) each form $C_2$ groups of order 2:

| $e$ | $a$ |
|---|---|
| $a$ | $e$ |

The real / imaginary pair, however, forms a $C_4$ group $(1, i, -1, -i)$ of order 4. The total group structure is then $C_4 \times C_2 \times C_2$ (or $C_4 \times D_2$), of order 16.

The 4-component structure does not, however, describe the complete symmetry of an object. Since we have three categories, each with two values, there is a total of eight parameters. In addition to the above four, there exist four dual parameters, given by

| **space\*** | imaginary | conserved | noncountable |
| **time\*** | real | conserved | countable |
| **mass\*** | imaginary | nonconserved | countable |
| **charge\*** | real | nonconserved | noncountable |

It has been suggested, previously, that these are the effective representations of space, time, mass and charge as used in the non-quantum-field version of the Dirac theory.[4-6] The 'dual group' to space, time, mass and charge will, of course, have many manifestations, of which the Dirac representation is just one. Such representations, will involve mathematical reversals of physical properties. However, it seems likely that any representation other than the canonical one (that is, the original group) will always be reduced to the canonical one under the process of measurement, as we have seen happens with the Heisenberg and Schrödinger formulations of quantum mechanics.

The dual parameters must possess the same $C_4 \times D_2$ structure, thus the total group symmetry must be of order 32. Additionally, these dual structures must mutually anticommute. Since the dual of $C_4$ is itself, the complete group structure must be $C_4 \times Q_4$, where $Q_4$ is the quaternion group (of order 8). This structure is simply the complexified quaternions. The total group structure may therefore be represented by a quaternion vector pair, where we have the following assignments:

| **Real scalar** | mass |
|---|---|
| **Imaginary scalar** | time |
| **Real 3-vector** | space |
| **Imaginary quaternion** | charge |

We see then that the symmetry structure of an object requires not only a 3-dimensional space, but also a 3-component charge. Futhermore, it is clear that mass and charge possesss the same general structure as space and time, coupled together to form a unified concept of 'mass-charge'.

### 6 Object symmetry and the Standard Model

Taken together, the complex quaternions above do not form a general group. It is clear, in fact, that general object symmetry cannot form a group, as the existence of nilpotents (such as these describe) is central to quantum duality. However, one is led to ask if the complex quaternion algebra is the product of some larger symmetry group, which appears in its current form via spontaneous symmetry breaking. There are a number of ways in which a supersymmetry can be created. Perhaps the simplest is to represent the state of an object as a vector-like term, as in

$$O => 1 + \mathbf{i} + \mathbf{j} + \mathbf{k} + i + i\mathbf{I} + i\mathbf{J} + i\mathbf{K}$$

In this form, the state of an object resembles a broken octonion. The dimensionality in this representation presupposes the concept of *rotation*, and the rotational symmetry of an octonion is the exceptional Lie group $G_2$. The above vector is then an octonion with one orientation (in this case $i$) constrained so as not to transform into other coordinates. The symmetry group of such a constrained octonion is $SU(3)$. Within this constrained symmetry, there exist two other obvious subgroups: the quaternion vector, with symmetry $SU(2)$, and the complex 'vector', with symmetry $U(1)$. It is clear then that object symmetry has the form $SU(3) \times SU(2) \times U(1)$, which can be generated from the spontaneous symmetry breaking of the octonion symmetry $G_2$.

It is clear the inherent symmetry of an abstract object, with no *a priori* internal structure is required to have the internal structure of the standard model. The standard model therefore arises from the basic properties of space, time, mass and charge.

## 7 Symmetry hierarchy

Although we have demonstrated the foundational approach is sufficiently general to encompass the standard model, we have not yet presented a complete picture of object symmetry. The foundational properties of space, time, mass, and charge mandate a direct connection to the full set of division algebras (real, complex, quaternion, octonion), as seen above. As a result, the foundational approach is closely related to a hierarchy of symmetry structures derived from these algebras. The most famous of these hierarchies is known as the Freudenthal-Tits Magic Square, which is a $4 \times 4$ array of groups, associated with the Jordan algebras of $3 \times 3$ Hermitian matrices. The result is derived from the groups $G \times G$, and is given by

|   | $R$ | $C$ | $Q$ | $O$ |
|---|------|------|------|------|
| $R$ | $SO(3) = SU(2)$ | $SU(3)$ | $Sp(3)$ | $F_4$ |
| $C$ | $SU(3)$ | $SU(3) \times SU(3)$ | $SU(6)$ | $E_6$ |
| $Q$ | $Sp(3)$ | $SU(6)$ | $SO(12)$ | $E_7$ |
| $O$ | $F_4$ | $E_6$ | $E_7$ | $E_8$ |

It is clear from this table that products of the octonion algebra, together with the octonion symmetry $G_2$, form the set of exceptional Lie algebras which are of such great interest in higher dimensional models such as string theory.

We have shown, then, a fundamental connection between the foundational symmetries of space, time, mass, and charge, and the 'higher' symmetries of models such as string theory and the standard model. Clearly, there exists enough freedom to express these models in a foundational context. Furthermore, one can argue that the foundational approach is more powerful, as it requires the above symmetries as a consequence, rather than imposing them *a priori*.

## 8 The $D2$ 'canonical' representation

The group representation that we choose for the fundamental parameters will depend on what we are actually looking at; purely 'qualitative' conceptions of a parameter's nature put into symbolic form will produce different representations to those derived using the internal algebras generated by the parameters themselves. In a purely symbolic sense, we may represent the properties of space (real, nonconserved, countable) by, say, $a$, $b$, $c$, with the opposing properties (imaginary, conserved, noncountable) represented by $-a$, $-b$, $-c$. The arrangement now becomes:

| | | | |
|---|---|---|---|
| **space** | $a$ | $b$ | $c$ |
| **time** | $-a$ | $b$ | $-c$ |
| **mass** | $a$ | $-b$ | $-c$ |
| **charge** | $-a$ | $-b$ | $c$ |

Mathematically, the scheme encompasses a group of order 4, in which any parameter can be the identity element and each is its own inverse. With multiplication rules of the form:

$$a * a = -a * -a = a$$
$$a * -a = -a * a = -a$$
$$a * b = a * -b = 0$$

and similarly for $b$ and $c$, we can establish a group multiplication table of the form:

| * | space | time | mass | charge |
|---|---|---|---|---|
| space | space | time | mass | charge |
| time | time | space | charge | mass |
| mass | mass | charge | space | time |
| charge | charge | mass | time | space |

This is the characteristic multiplication table of the Klein-4 or $D_2$ group, with space as the identity element and each element its own inverse. However, there is no reason to privilege space with respect to the other parameters, since the symbols $a$ and $-a$, $b$ and $-b$, $c$ and $-c$ are arbitrarily selected, and any of the other three parameters may be made the identity by defining its properties as $a$, $b$, $c$. For example, if mass is made the identity element, then the group properties may be represented by:

| | | | |
|---|---|---|---|
| **space** | $a$ | $-b$ | $-c$ |
| **time** | $-a$ | $-b$ | $c$ |
| **mass** | $a$ | $b$ | $c$ |
| **charge** | $-a$ | $b$ | $-c$ |

and the multiplication table becomes:

| * | mass | charge | time | space |
|---|---|---|---|---|
| mass | mass | time | charge | space |
| charge | time | mass | space | charge |
| time | charge | space | mass | time |
| space | space | time | charge | mass |

The dual group may be represented symbolically by:

| | | | |
|---|---|---|---|
| **space**\* | $-a$ | $b$ | $c$ |
| **time**\* | $a$ | $b$ | $-c$ |
| **mass**\* | $-a$ | $-b$ | $-c$ |
| **charge**\* | $-a$ | $-b$ | $c$ |

In this case the multiplication rule is:

$$a * a = -a * -a = -a$$
$$a * -a = -a * a = a$$
$$a * b = a * -b = 0$$

and the group multiplication table becomes:

| *       | space*  | time*   | mass*   | charge* |
|---------|---------|---------|---------|---------|
| space*  | mass*   | charge* | space*  | time*   |
| time*   | charge* | mass*   | time*   | space*  |
| mass*   | space*  | time*   | mass*   | charge* |
| charge* | time*   | space*  | charge* | mass*   |

Here, mass* becomes the identity element, though, again this arbitrary, and changing the signs of both $b$ and $c$, for example, would make space* the identity element.

That $D_2$ is more fundamental than $C_2$ may seem at first a surprising result, as $C_2$ is the simplest possible group, a direct description of the 'something from nothing'-type duality with which we began. However, physics is ultimately structured on the need to define a concept of *measurement* (the 'probe') and a single fundamental $C_2$ would not allow this. By definition, a perfect $C_2$ symmetry would simply create an unrecognizable 'response' to a measurement probe; and, while a perfect $C_2$ would be meaningless, an 'imperfect' $C_2$ (allowing oppositeness in only a limited number of characteristics) would be incomplete. The simplest perfect symmetry (in the physical sense) based on the $C_2$ principle, which allows recognition of the response, is then $D_2$.

## 9 Conservation laws and fundamental symmetries

Assuming that the $D_2$ group containing space, time, mass and charge is the only available physical information at the fundamental level, it is possible to provide derivations of the equations of classical mechanics and electromagnetic theory.[4-6] The already existing numerical relations between space and time units, and mass and charge units, extend also by group symmetry to relations between the units of every fundamental quantity and every other, and their inverses. This is the origin of $G$, $h$ and $c$ as fundamental constants. Quantum mechanics follows when we define a new conserved operator and apply total variation of space and time. At the same time, the continuity of mass-energy defines a new meaning for the field equations of general relativity, directly incorporating inertial effects while preserving the classical structure of gravity.[12-16] However, completely new mathematical results can also be generated by even more direct uses of the symmetries. These results are extensions of the application of Noether's theorem to the conservation laws of energy, momentum and angular momentum. In fact, Noether's theorem itself, linking conserved quantities with transformations, is really an expression of the fact that every fundamental conserved quantity must be symmetrical to a fundamental nonconserved quantity.

In one application of Noether's theorem, as we have seen, we link the conservation of energy (or mass, in our terms) to the translation symmetry of time. In effect, the nonconservation of time is responsible for the conservation of mass, a result which is evident in our original symmetry group. By extending the analogy, we can link the conservation of the quantity of charge with the nonconservation, or translation symmetry of space, and, hence, the conservation of linear momentum.[4-6] We can, therefore, propose a theorem in which the conservation of linear momentum is responsible for the conservation of the quantity of charge – of any type. We can also link the conservation of *type* of charge to the rotation symmetry of space, or the conservation of angular momentum, as in the following scheme:

| symmetry | conserved quantity | linked conservation |
|---|---|---|
| space translation | linear momentum | value of charge |
| time translation | energy | value of mass |
| space rotation | angular momentum | type of charge |

In fact, physical theorems already exist which show these principles to be true in special cases. The conservation of electric charge within a system was shown by Fritz London, in 1927, to be identical to invariance under transformations of the electrostatic potential by a constant representing changes of phase, with the phase changes being of the kind involved in the conservation of linear momentum. In a conservative system, electrostatic potential varies only with the spatial coordinates, and so the London principle implies that the quantity of electric charge is conserved *because* the spatial coordinates are not. This is, therefore, a special case of the first predicted relation.

In the second case, we have the relation between spin and statistics observed in fundamental particles. Fermions have half-integral spin angular momentum, while for bosons this quantity becomes integral; fermions also have weak units of charge, while bosons have none. The presence of a particular type of charge thus appears to determines the angular momentum state of the particle, so conservation of this type of charge in a particle is linked with the conservation of its angular momentum state. We will subsequently show that the general theorem is fundamental to the understanding of particle structures and to the symmetry-breaking between the electromagnetic, weak and strong interactions.

## 10 A hierarchy of dualities

We can, as we have seen, construct a $D_2$ group for space, time, mass and charge on the basis of assigning algebraic symbols to properties without concerning ourselves about the actual mathematics associated with the four parameters. However, as soon as one assigns such mathematics, the group relationship changes. The resulting structures form a hierarchy of dualities, analogous to the hierarchy of dualities that we can use to create the natural numbers in binary form. The hierarchy stems from the fact that any mathematical representation of a symmetry will necessarily be an imperfect or 'broken' symmetry, like the definition of a 'system' in the context of Newton's third law of motion, and will necessarily lead to a doubling process, tending to a more 'perfect' symmetry.

The doubling can be seen as a realization of the fact that everything has to have a dual or $C_2$ partner to make something from nothing, and the mathematics within the parameter structures has to be constructed so that the doubling process is actually possible. The real / imaginary distinction and 3-dimensionality are ways in which doubling can occur, but they act in subtly different ways. Both ideas are associated with the squaring process, which is a kind of doubling, just as square rooting is a kind of halving. This means that squaring via $C_2$ is as natural a way of producing 'something from nothing'-type duality as adding.

We can explain the duality hierarchy in terms of a 'Fundamental Theorem of Symmetry': For every symmetry, there must exist a corresponding asymmetry. As such, a simple duality can always be represented as $C_2$. If we define science as the measurement of a system to determine results, there will be three basic duals. The System Dual corresponds to conservation / nonconservation. A system consists of an object within an environment. The object and the environment are therefore dual. Mass and charge, here, are object properties, space and time are environment properties. The Measurement Dual corresponds to countability / noncountability (discreteness / continuity). For any system, counting is either possible, or it is not. Thus 'measurable' and 'nonmeasurable' (countable and noncountable) are dual. Space and charge are countable, time and mass are not. The Observation Dual does not correspond directly to the real / imaginary distinction, but it creates it indirectly in the form of the duality between the 'canonical' $D_2$ group and its own dual, and it is central to the idea of 'probe' and 'response'. Any measurement consists of the observer and the observed. Thus observer / observed are dual.

Though these three dualities must exist within any scientific model, they also generate further dualities. The System / Observation also forms a duality, since it creates two possibilities: the object is the observer and the environment is the observed, or the environment is the observer and the object is the observed. It is this duality which is expressed when we reverse the physically incorrect assumptions of either the Schrödinger or the Heisenberg mathematical systems in the process of measurement.

The System / Observation Dual corresponds directly to the real / imaginary distinction since it is a two-fold dual. That is, there are four quantities (object-observer, object-observed, environment-observer, environment-observed), and they must have an internal duality. This is why the real / imaginary distinction becomes $C_4 = (1, -1, i, -i)$, where $(1, -1)$ and $(i, -i)$ have a dual symmetry. This last duality is also what generates the classical / quantum symmetry, and can explain why the $i$ shows up in quantum mechanics. We can also say that the real / imaginary distinction is what doubles the original group to include the dual group. So the creation of a dual group is the same thing, in effect, as making the real / imaginary distinction explicit. We can further suppose that any of the $C_2$ dualities in space, time, mass and charge will produce the factor 2, which occurs in many aspects of physics as a direct mathematical expression of the fundamental nature of duality, and that one duality can be translated into another. The $C_2$ 'something from nothing' doubling is then extended to higher orders automatically, because of its own 'self-dynamism'.

The important aspect for the space, time, mass, charge model is that there is an asymmetry in the formulation. In this sense, even the fundamental parameter group of space, time, mass and charge is a *mathematically* (though not physically) broken symmetry, because the full $C_2 \times C_2 \times C_2$ (which includes the dual group) is not needed for physics at any given time, but only half of it. The System, Observation, and System-Observation duals, also, form a closed set. The Measurement duality is not part of this set, rather it is a kind of truth 'external' to the system. This means the four duals do not form a group, but require a set of order 32 which is not a group, but which has the complex numbers as a subgroup. Thus, the result is the complexified quaternions. The higher doublings then occur through the recognition of this structure as a larger group structure, and lead, through the application of the rotation symmetries implicit in the multidimensional parameters, to the creation of the Lie algebras.

## 11 Dimensionality

We have seen that, bringing in the real / imaginary distinction doubles the order of the group; introducing 3-dimensionality doubles it again. The doubling effect is natural, due to the basic concept of duality, and the 3-dimensionality is itself related to the real / imaginary distinction, but the process of creating multidimensionality by doubling is very subtle. The only way for a group to have some form of 'dimensionality' is for it to be non-abelian or non-commutative in some form. It is not possible for a non-abelian group to be the product of abelian groups. Therefore, one cannot simply 'generate' dimensionality out of the 'doubling' effect, alone. The smallest dimensional finite group is the quaternion group, which is of dimension 3 (not 4, since 'time' is commutative). This appears at order 8. Of the 5 finite groups of order 8, $C_8$, $C_2 \times C_4$, $C_2 \times C_2 \times C_2$, $D_4$, and $Q$, all are abelian except for $Q$. So, somehow there must be an argument requiring $Q$, or at some higher order, a non-abelian group of which $Q$ is a subgroup.

3-dimensionality is a source of the factor 2 in many aspects of physics. Dynamically and quantum mechanically, the factor 2 is associated with vector terms. Dynamically it comes from action and reaction, or the virial theorem (as a direct product of 3-dimensionality); quantum mechanically, it comes from the noncommutativity of the vector terms in the spin angular momentum. In general, the factor 2 in physics comes from the division between discreteness – the source of multidimensionality – and continuity. In the group structures it is also associated with the division between real and complex representations. The complex version of Dirac algebra (as will be described in section 13) has 32 terms, as opposed to the 'real' version with 16. It must be relevant, therefore, that it is the imaginary version of vector algebra (in the quaternion) which requires 3-dimensionality or 3+1-dimensionality.

In principle, the 'rule of doubling' means that the multidimensionality of quaternions has to be of such an order as would create the *same number of elements* as we would get by doubling, even though they will not be the *same elements*. That is, considerations of duality mean that we need a doubling of elements, but *some other fundamental consideration* requires that the perfection of the originally abelian symmetry is violated, and the doubling occurs in a 'broken' way. Duality itself creates the spatial 3-dimensionality, and the 3+1 dimensionality of, say, space plus time, mass and charge (or space, time and mass plus charge) becomes the double 3+1 dimensionality of space plus time and mass plus charge when we double the grouping by including the dual group. But this is a requirement, not a *prescription*.

In the space, time, mass, charge set, 3-dimensionality becomes a necessary result, through symmetry, of extending the real / imaginary distinction from time to charge. Charge, being discrete, is necessarily multidimensional, and, being also imaginary, requires a description in the form of a multidimensional imaginary algebra. What seemingly happens is that the group duality consideration, requiring a multiplication of elements such that the number of elements increases from 4 to 8, but allowing only 2 of the elements to acquire the property that would make this possible, requires 3-dimensionality, and this 3-dimensionality is expressed by what we happen to call quaternions, of order 8, which are noncommutative because of the rotation property. The real / imaginary division itself then requires a further multiplication of elements from 8 to 16, creating, in the process, the parallel 3-dimensionality of the real vector parameter, space.

One method of producing the quaternion group *mathematically* would be to start with the three dualities, of conservation / nonconservation, countability / noncountability, real / imaginary, and treat them initially as simple $C_2$ symmetries, with two categories, exactly opposite. These product to $C_2 \times C_2 \times C_2$, of order 8. But then, if we take the real / imaginary duality, although only contributing an order 2, as, mathematically, of the form $C_4$, we could, by symmetry, take each of the dualities to be of the same form. In other words, the group of order 8 would have 3 unique $C_4$ subgroups. $Q$ has three $C_4$ subgroups as its only subgroups. Thus, the order 8 group would be $Q$, automatically introducing dimensionality of order 3.

Like the direct doubling due to the real / imaginary distinction (which creates mathematics of ordered pairs), it would seem that the dimensional doubling is related to the introduction of imaginary numbers. Both are a product of the Pythagorean squaring process, which is simultaneously an expression of dimensionality and of the group relationships between the fundamental parameters, in addition to being the actual mechanism by which the group doubling process becomes related to the mathematical factor 2. In the case of multidimensional quantities this factor is derived by making dimensional multiplication noncommutative, and only 3-dimensional quantities may accomplish this in a way that preserves symmetrical closure. In addition, symmetry considerations determine that only two parameters may become multidimensional, and so only two have this property of noncommutativity.

From a purely *physical* point of view, the central origin of dimension can be taken as the particle / wave duality, which comes directly from the discrete / continuous division, for this is certainly the origin of the division between multidimensionality and unidimensionality. Only discrete quantities can have more than one dimension, and discreteness requires more than one dimension; while continuous quantities must necessarily be unidimensional. Particles express discreteness, though they have no dimension (or extension) themselves; waves express continuity, but, as the opposite of particles, must, in a sense, have dimension. Effectively, particles mark the discreteness of space, by being the divisions or zero points of it, and so, in this sense, are not discrete at all. Once again, this is similar to the Heisenberg-Schrödinger distinction. Each incorporates its opposite to be able to define itself. There is a dual aspect to everything.

## 12 A broken octonion

We can consider the original group, space, time, mass and charge and dual group, space*, time*, mass*, charge*, as the 'commutative' and 'noncommutative' ways of producing the abstract $D_2$ structure, and we can combine the two into a larger group structure $C_2 \times D_2$ of order 8. For example, the identity element, say mass, could be represented by the scalar part of a quaternion (1) and the other three terms by the imaginary operators $i, j, k$, while mass*, charge*, space*, time* are represented by $-1, -i, -j, -k$. So, the multiplication table:

| *    | 1    | $i$  | $j$  | $k$  | $-1$ | $-i$ | $-j$ | $-k$ |
|------|------|------|------|------|------|------|------|------|
| 1    | 1    | $i$  | $j$  | $k$  | $-1$ | $-i$ | $-j$ | $-k$ |
| $i$  | $i$  | $-1$ | $k$  | $-j$ | $-i$ | 1    | $-k$ | $j$  |
| $j$  | $j$  | $-k$ | $-1$ | $i$  | $-j$ | $k$  | 1    | $-i$ |
| $k$  | $k$  | $j$  | $-i$ | $-1$ | $-k$ | $-j$ | $i$  | 1    |
| $-1$ | $-1$ | $-i$ | $-j$ | $-k$ | 1    | $i$  | $j$  | $k$  |
| $-i$ | $-i$ | 1    | $-k$ | $j$  | $i$  | $-1$ | $k$  | $-j$ |
| $-j$ | $-j$ | $k$  | 1    | $-i$ | $j$  | $-k$ | $-1$ | $i$  |
| $-k$ | $-k$ | $-j$ | $i$  | 1    | $k$  | $j$  | $-i$ | $-1$ |

can be used to represent the group relations between M, C, S, T and M*, C*, S*, T*:

| * | M | C | S | T | M* | C* | S* | T* |
|---|---|---|---|---|----|----|----|----|
| M | M | C | S | T | M* | C* | S* | T* |
| C | C | M* | T | S* | C* | M | T* | S |
| S | S | T* | M* | C | S* | T | M | C* |
| T | T | S | C* | M* | T* | S* | C | M |
| M* | M* | C* | S* | T* | M | C | S | T |
| C* | C* | M | T* | S | C | M* | T | S* |
| S* | S* | T | M | C* | S | T* | M* | C |
| T* | T* | S* | C | M | T | S | C* | M* |

If the 3-dimensionality of charge and space is directly involved, the overall structure requires a quaternion ($\pm 1$, $\pm \boldsymbol{i}s$, $\pm \boldsymbol{j}e$, $\pm \boldsymbol{k}w$) and a quaternion-like 4-vector ($\pm i$, $\pm \boldsymbol{i}x$, $\pm \boldsymbol{j}y$, $\pm \boldsymbol{k}z$) (a double quaternion in total) within another overall quaternion-type arrangement. This can be accomplished using an octonion, with sixteen members ($\pm 1m$, $\pm \boldsymbol{i}s$, $\pm \boldsymbol{j}e$, $\pm \boldsymbol{k}w$, $\pm \boldsymbol{e}t$, $\pm \boldsymbol{f}x$, $\pm \boldsymbol{g}y$, $\pm \boldsymbol{h}z$) (see below). Although this is no longer a group, it combines two conjugate groups of order 8: M, C(3), S(3), T and M*, C(3)*, S(3)*, T*, and can be represented in a group structure through the use of left-product or right-product octonions. In addition, the nonassociativity of the dimensional terms in this octonion extension seems to be lost within terms which effectively cancel each other out, and are of no physical significance.

| * | 1 | i | j | k | e | f | g | h |
|---|---|---|---|---|---|---|---|---|
| 1 | 1 | i | j | k | e | f | g | h |
| i | i | −1 | k | −j | f | −e | −h | g |
| j | j | −k | −1 | i | g | h | −e | −f |
| k | k | j | −i | −1 | h | −g | f | −e |
| e | e | −f | −g | −h | −1 | i | j | k |
| f | f | e | −h | g | −i | −1 | −k | j |
| g | g | h | e | −f | −j | k | −1 | −i |
| h | h | −g | f | e | −k | −j | i | −1 |

| * | m | s | e | w | t | x | y | z |
|---|---|---|---|---|---|---|---|---|
| m | m | s | e | w | t | x | y | z |
| s | s | −m | w | −e | x | t | −z | y |
| e | e | −w | −m | s | y | z | −t | −x |
| w | w | e | −s | −m | z | −y | x | −t |
| t | t | −x | −y | −z | −m | s | e | w |
| x | x | t | −z | y | −s | −m | −w | e |
| y | y | z | t | −x | −e | w | −m | −s |
| z | z | −y | x | t | −w | −e | s | −m |

If we take charge as the identity element, and represent it by a scalar, the remaining structure for time, space and mass (and, implicitly, the energy, momentum and mass operators) becomes that of the Dirac algebra, and $SU(5)$ or $U(5)$ (as described later). Such representations do not determine the properties of the group members, space, time, mass and charge. They exist because the group has four components, and can, therefore, be represented by a 4-component structure like a quaternion, in which the link between elements is made by a binary operation (squaring); but the link between a group with four components and a 4-dimensional space-time or mass-charge may be in itself significant. In addition, the $D_2$ group is the group of rotations of the rectangle (identity and rotations about three spatial axes). This could be another way of linking the double 3+1 symmetry of the units of space, time, mass and charge with the structure of the group.

## 13 The Dirac algebra

Only two members of the original group, say space and time, are needed to define the structure. We can create quantum mechanics with a third, say mass or charge. A broken octonion symmetry creates for us the algebra we will need for quantum mechanics. Here, the 8-unit octonion splits into a 4-vector part, representing space and time, with real vector units **i**, **j**, **k** and imaginary scalar $i$; and a quaternion part, representing charge and mass, with imaginary vector units *i*, *j*, *k* and real scalar 1. The quaternions follow the usual multiplication rules:

$$i^2 = j^2 = k^2 = ijk = -1$$
$$ij = -ji = k$$
$$jk = -kj = i$$
$$ki = -ik = j ,$$

while, to make the two algebras completely symmetrical, the real vector units follow multivariate multiplication rules, identical to those for Pauli matrices, and parallel to those for quaternion algebra:

$$\mathbf{i}^2 = \mathbf{j}^2 = \mathbf{k}^2 = 1$$
$$\mathbf{ij} = -\mathbf{ji} = i\mathbf{k}$$
$$\mathbf{jk} = -\mathbf{kj} = i\mathbf{i}$$
$$\mathbf{ki} = -\mathbf{ik} = i\mathbf{j} \ .$$

In effect, this requires a 'full product' for two vectors **a** and **b** of the form

$$\mathbf{ab} = \mathbf{a.b} + i \ \mathbf{a} \times \mathbf{b} \ .$$

Recombining the two algebras using the eight base units, $\mathbf{i}$, $\mathbf{j}$, $\mathbf{k}$, $i$, $i$, $j$, $k$, 1, we define a new algebra with 32 parts, which is isomorphic to the Dirac algebra, defined by the 5 $\gamma$ matrices,[17-25] and we can make correlations of the form:

$$\gamma^0 = -i\mathbf{i} \qquad \text{or} \qquad \gamma^0 = i\mathbf{k}$$
$$\gamma^1 = \mathbf{ik} \qquad\qquad\qquad \gamma^1 = \mathbf{ii}$$
$$\gamma^2 = \mathbf{jk} \qquad\qquad\qquad \gamma^2 = \mathbf{ji}$$
$$\gamma^3 = \mathbf{kk} \qquad\qquad\qquad \gamma^3 = \mathbf{ki}$$
$$\gamma^5 = ij \qquad\qquad\qquad \gamma^5 = ij \ .$$

The complete algebra incorporates: 1 real scalar, 1 imaginary scalar, 3 real vectors, 3 imaginary vectors, 3 quaternions, 3 imaginary quaternions, 9 real vector quaternions and 9 imaginary vector quaternions. The full 32 parts can be derived as follows:

$\gamma^0 = i\mathbf{k}$, $\gamma^1 = \mathbf{ii}$, $\gamma^2 = i\mathbf{j}$, $\gamma^3 = i\mathbf{k}$, $\gamma^5 = ij$,

$\gamma^0\gamma^1 = ij\mathbf{i}$, $\gamma^0\gamma^2 = ij\mathbf{j}$, $\gamma^0\gamma^3 = ij\mathbf{k}$, $\gamma^0\gamma^5 = i$, $\gamma^1\gamma^2 = -i\mathbf{k}$,
$\gamma^1\gamma^3 = i\mathbf{j}$, $\gamma^1\gamma^5 = i k\mathbf{i}$, $\gamma^2\gamma^3 = -i\mathbf{i}$, $\gamma^2\gamma^5 = ik\mathbf{j}$, $\gamma^3\gamma^5 = ik\mathbf{k}$,

$\gamma^0\gamma^1\gamma^2 = k\mathbf{k}$, $\gamma^0\gamma^1\gamma^3 = -k\mathbf{j}$, $\gamma^0\gamma^1\gamma^5 = \mathbf{i}$, $\gamma^0\gamma^2\gamma^3 = ki\mathbf{i},\gamma^0\gamma^2\gamma^5 = \mathbf{j}$,
$\gamma^0\gamma^3\gamma^5 = \mathbf{k}$, $\gamma^1\gamma^2\gamma^3 = -ii$, $\gamma^1\gamma^2\gamma^5 = j\mathbf{k}$, $\gamma^1\gamma^3\gamma^5 = -jj$, $\gamma^2\gamma^3\gamma^5 = ji$,

$\gamma^0\gamma^1\gamma^2\gamma^3 = j$, $\gamma^0\gamma^1\gamma^2\gamma^5 = -ii\mathbf{k}$, $\gamma^0\gamma^1\gamma^3\gamma^5 = ii\mathbf{j}$, $\gamma^0\gamma^2\gamma^3\gamma^5 = -ii\mathbf{i}$, $\gamma^1\gamma^2\gamma^3\gamma^5 = k$,

$\gamma^0\gamma^1\gamma^2\gamma^3\gamma^5 = -i$.

The 32 parts become a group of order 64 when + and − values of the terms are taken into account. Terms of the opposite sign are produced by reversing the order of multiplication.

## 14 The nilpotent formulation of quantum mechanics

Using this 32-part algebra, we can easily derive the Dirac equation from the relativistic momentum-energy conservation equation

$$E^2 - p^2 - m^2 = 0 \ .$$

We factorize and attach the exponential term $e^{-i(Et - \mathbf{p.r})}$ (which expresses the required total variation in space and time), so that

$$(\pm \boldsymbol{k}E \pm \boldsymbol{ii}\ \mathbf{p} + \boldsymbol{ij}\ m)\,(\pm \boldsymbol{k}E \pm \boldsymbol{ii}\ \mathbf{p} + \boldsymbol{ij}\ m)\ e^{-i(Et - \mathbf{p.r})} = 0 \ ,$$

and then replace $E$ and $\mathbf{p}$ in the first bracket with the quantum operators, $i\partial / \partial t$ and $-i\nabla$, to give

$$\left( \pm i\boldsymbol{k}\frac{\partial}{\partial t} \pm i\boldsymbol{\nabla} + \boldsymbol{ij}m \right)(\pm \boldsymbol{k}E \pm \boldsymbol{ii}\ \mathbf{p} + \boldsymbol{ij}\ m)\ e^{-i(Et - \mathbf{p.r})} = 0 \ .$$

This can be written in the form

$$\left( \pm i\boldsymbol{k}\frac{\partial}{\partial t} \pm i\boldsymbol{\nabla} + \boldsymbol{ij}m \right)\psi\ =\ 0 \ ,$$

where the wavefunction,

$$\psi = (\pm \boldsymbol{k}E \pm \boldsymbol{ii}\ \mathbf{p} + \boldsymbol{ij}\ m)\ e^{-i(Et - \mathbf{p.r})} \ \ .$$

The vector elements associated with the second term are, of course, multivariate, and we are free to replace $\mathbf{p}$ with terms like $\mathbf{1.p}$ or $\boldsymbol{\sigma}.\mathbf{p}$ (the spin term derivable directly from the equation itself). With the $m$ term fixed as positive, the equation defines four solutions, which represent the four possible combinations of $\pm E$ (particle / antiparticle) and $\pm \mathbf{p}$ or $\boldsymbol{\sigma}.\mathbf{p}$ (spin up / down). These four $(= 2^{D/2})$ solutions are demanded by the use of a quaternion representation and a 4-D space-time, and are most conveniently represented using a column (or row) vector, as in the standard representation of the Dirac spinor. Superpositions of two states will then require a scalar product between the four terms representing each state.

The wavefunctions are nilpotents or square roots of zero, and the four terms of the form $(\pm \boldsymbol{k}E \pm \boldsymbol{ii}\ \mathbf{p} + \boldsymbol{ij}\ m)$ are anticommuting pentads, with five elements, including three for the vector term $\boldsymbol{ii}\ \mathbf{p}$. The nilpotent aspect is particularly significant, as we have seen, in the context of topos theory, and in the general context of deriving something from nothing. Without the anticommuting pentad arrangement, involving a complexified vector-quaternion algebra, this would be impossible.

The nilpotent wavefunctions have the advantage of automatically introducing second quantization, because both the differential operator and the wavefunction are quantized to the same degree. Second quantization is a reflection of the duality in the underlying group structure. (It also reflects the fact that the nilpotent wavefunction incorporates both 'Heisenberg' and 'Schrödinger' elements, in the respective terms ($\pm$ *$kE$* $\pm$ *$ii$* **p** + *$ij$* *$m$*) and $e^{-i(Et - \mathbf{p.r})}$.) Pauli exclusion is an immediate consequence of the formulation, as the product of two identical fermion wavefunctions is necessarily zero. For the same reason, nilpotency automatically leads to nonlocal correlation between all fermion states (because all must 'know' immediately that none is identical to any other), at the same time as it requires time-delayed local interactions between particle states by being the square root of a relativistic energy conservation equation.

The scalar product of two identical states, in the nilpotent formulation, is necessarily zero. If, however, we reverse the signs of either *$kE$* or *$ii$* **p** or both in one of the two states, we will immediately produce a nonzero scalar product, such as a multiple of $E^2 + p^2 - m^2$, or $-E^2 + p^2 - m^2$, when we sum up over the four solutions representing the states. A vector boson (spin = 1) has fermion and antifermion components with the same sign of **p**, but the opposite sign of $E$. So its wavefunction incorporates a scalar product of the form:

$$(kE + ii\,\mathbf{p} + ij\,m)\,(-kE + ii\,\mathbf{p} + ij\,m)\ ;$$
$$(kE - ii\,\mathbf{p} + ij\,m)\,(-kE - ii\,\mathbf{p} + ij\,m)\ ;$$
$$(-kE + ii\,\mathbf{p} + ij\,m)\,(kE + ii\,\mathbf{p} + ij\,m)\ ;$$
$$(-kE - ii\,\mathbf{p} + ij\,m)\,(kE - ii\,\mathbf{p} + ij\,m)\ .$$

whose sum is a nonzero scalar, $4\ (E^2 + p^2 + m^2) = 8\ E^2$, before normalization. A scalar boson (spin = 0), however, has components with **p** terms of opposite signs. Hence its wavefunction incorporates a scalar product of the form:

$$(kE + ii\,\mathbf{p} + ij\,m)\,(-kE - ii\,\mathbf{p} + ij\,m)\ ;$$
$$(kE - ii\,\mathbf{p} + ij\,m)\,(-kE + ii\,\mathbf{p} + ij\,m)\ ;$$
$$(-kE + ii\,\mathbf{p} + ij\,m)\,(kE - ii\,\mathbf{p} + ij\,m)\ ;$$
$$(-kE - ii\,\mathbf{p} + ij\,m)\,(kE + ii\,\mathbf{p} + ij\,m)\ ,$$

which is, again, a scalar, $4\ (E^2 - p^2 + m^2) = 8\ m^2$, before normalization, while a Bose-Einstein condensate wavefunction incorporates a scalar product which is the sum of:

$$(kE + ii\,\mathbf{p} + ij\,m)\,(kE - ii\,\mathbf{p} + ij\,m)\ ;$$
$$(kE - ii\,\mathbf{p} + ij\,m)\,(kE + ii\,\mathbf{p} + ij\,m)\ ;$$
$$(-kE + ii\,\mathbf{p} + ij\,m)\,(-kE - ii\,\mathbf{p} + ij\,m)\ ;$$
$$(-kE - ii\,\mathbf{p} + ij\,m)\,(-kE + ii\,\mathbf{p} + ij\,m)\ ,$$

which is the scalar $4\ (-E^2 - p^2 + m^2) = -8\ p^2$, before normalization. In the case of correlations between wavefunctions of noninteracting fermions, say, ($kE_1 + ii\ \mathbf{p}_1 + ij\ m_1$), ($kE_2 + ii\ \mathbf{p}_2 + ij\ m_2$), etc, we clearly obtain a bosonic state for an even number of wavefunctions, and a fermionic state for an odd number.

The nilpotent operators are also creation and annihilation operators and equivalent to the quantum field integrals acting on the vacuum state. The $Q$ generator for supersymmetry, converting bosons to fermions, is simply the term ($kE + ii\mathbf{p} + ijm$), while $Q\dagger = (-kE + ii\mathbf{p} + ijm$) converts bosons to antifermions, or fermions to bosons, assuming, of course, that the full operators are four-term bra or ket vectors, with the $E$ and $\mathbf{p}$ values going through the usual cycle of $+$ and $-$ values. The nilpotent formulation shows clearly the relationship between spin and statistics between the particle types. The fermion wavefunction is a single nilpotent, a square-root of the zero squared-energy term. The spin ½ term comes immediately from the square-rooting operation. The boson wavefunction, on the other hand, is a nonzero scalar product of two nilpotents, and combines two spin ½ terms into an integral multiple. The square-rooting introduces the noncommutativity of the wavefunction and Pauli exclusion at the same time as it introduces the factor ½ into the spin. In fact, it uses the same mathematical procedure.

Square-rooting the energy equation also introduces discreteness (the fermion, with rest mass) with a continuous variation (the kinetic energy). The factor 2 which appears so frequently in fundamental physics can nearly always be expressed in terms of a transition between discreteness and continuity, and is, in this sense, a direct expression of the duality of the group structure.[10] The other dualities (real / imaginary, conserved / nonconserved) have the same effect – complexification, for example, introduces discreteness in the creation of the Dirac pentad. The factor 2 is also an expression of dimensionality, as can be seen from the fact that it arises in the square rooting of the energy-squared term. It is significant that complex and quaternionic factors only appear in the linear form, and never in the squared form used in Pythagorean-type addition.

The fact that $\mathbf{p}$ is a 3-component quantity means that it is possible to have a fermion-like 3-component nilpotent wavefunction representing a single quantum mechanical (baryon) system; such possibilities also derive directly from the pentad algebra. Baryon wavefunctions constructed from nilpotents use the three-dimensional properties of the $\mathbf{p}$ operator, by representing the six degrees of freedom for the spin as $\pm\ p_1$, $\pm\ p_2$, $\pm\ p_3$, and assuming that each one of these represents a phase of the interaction between the three *quark* components. Then an expression of the form

$$(kE \pm ii\ p_1 + ij\ m)\,(kE \pm ii\ p_2 + ij\ m)\,(kE \pm ii\ p_3 + ij\ m)\,,$$

by successively taking $\mathbf{p}$ through each of the phases, incorporates all six terms in the antisymmetric colour singlet of $SU(3)$:

$$\psi \sim (BGR - BRG + GRB - GBR + RBG - RGB)\,.$$

A vacuum operator may also be constructed from nilpotents in the form of a diagonal matrix, premultiplied by a 4-component row state vector or postmultiplied by a 4-component column state vector, representing a fermion state. In the first case, we write:

$$( (-kE - i\mathbf{i}\mathbf{p} + i\mathbf{j}m) \ (-kE + i\mathbf{i}\mathbf{p} + i\mathbf{j}m) \ (kE - i\mathbf{i}\mathbf{p} + i\mathbf{j}m) \ (kE + i\mathbf{i}\mathbf{p} + i\mathbf{j}m)) \times$$

$$\mathbf{k} \begin{pmatrix} kE + i\mathbf{i}\mathbf{p} + i\mathbf{j}m & 0 & 0 & 0 \\ 0 & kE - i\mathbf{i}\mathbf{p} + i\mathbf{j}m & 0 & 0 \\ 0 & 0 & -kE + i\mathbf{i}\mathbf{p} + i\mathbf{j}m & 0 \\ 0 & 0 & 0 & -kE - i\mathbf{i}\mathbf{p} + i\mathbf{j}m \end{pmatrix} e^{-i(Et - \mathbf{p} \cdot \mathbf{r})}$$

$$= ( (-kE - i\mathbf{i}\mathbf{p} + i\mathbf{j}m) \ (-kE + i\mathbf{i}\mathbf{p} + i\mathbf{j}m) \ (kE - i\mathbf{i}\mathbf{p} + i\mathbf{j}m) \ (kE + i\mathbf{i}\mathbf{p} + i\mathbf{j}m)) \, e^{-i(Et - \mathbf{p} \cdot \mathbf{r})} ,$$

assuming the appropriate normalisation constants. When applied to a row vector, the vacuum operator is always $\mathbf{k} \times$ matrix form of state vector for the fermion. The order is reversed when applied to a column vector.

For a single fermion interacting with the vacuum, we can imagine an infinite series of terms of the form $(kE + i\mathbf{i}\mathbf{p} + i\mathbf{j}m)$, $(kE + i\mathbf{i}\mathbf{p} + i\mathbf{j}m)(-kE + i\mathbf{i}\mathbf{p} + i\mathbf{j}m)$, $(kE + i\mathbf{i}k\mathbf{p} + i\mathbf{j}m)(-kE + i\mathbf{i}\mathbf{p} + i\mathbf{j}m)(kE + i\mathbf{i}\mathbf{p} + i\mathbf{j}m)$, $(kE + i\mathbf{i}\mathbf{p} + i\mathbf{j}m)(-kE + i\mathbf{i}\mathbf{p} + i\mathbf{j}m)(kE + i\mathbf{i}\mathbf{p} + i\mathbf{j}m)(-kE + i\mathbf{i}\mathbf{p} + i\mathbf{j}m)$, etc., allowing a cancellation of the boson and fermion loops of opposite sign in a kind of natural supersymmetry without additional particles. To convert bosons to fermions we multiply by $(kE + i\mathbf{i}\mathbf{p} + i\mathbf{j}m)$. To convert bosons to antifermions, or fermions to bosons, we multiply by $(-kE + i\mathbf{i}\mathbf{p} + i\mathbf{j}m)$. In addition, $(kE + i\mathbf{i}\mathbf{p} + i\mathbf{j}m) \ (-kE + i\mathbf{i}\mathbf{p} + i\mathbf{j}m) \ (kE + i\mathbf{i}\mathbf{p} + i\mathbf{j}m)( -kE + i\mathbf{i}\mathbf{p} + i\mathbf{j}m)$ ... is the same as $(kE + i\mathbf{i}\mathbf{p} + i\mathbf{j}m) \, \mathbf{k} \ (kE + i\mathbf{i}\mathbf{p} + i\mathbf{j}m) \, \mathbf{k} \ (kE + i\mathbf{i}\mathbf{p} + i\mathbf{j}m) \, \mathbf{k} \ (kE + i\mathbf{i}\mathbf{p} + i\mathbf{j}m)$ ... , which expresses the repeated action of the fermion state $(kE + i\mathbf{i}\mathbf{p} + i\mathbf{j}m)$ on the vacuum state $\mathbf{k} \ (kE + i\mathbf{i}\mathbf{p} + i\mathbf{j}m)$. The nilpotent representation of point fermions, creating an infinite series of further nilpotents in the continuous vacuum, suggests a connection with topos theory, as we have previously stated. In this connection, it may be of significance that the relativistic time-space-proper time combination can also be considered as a Dirac nilpotent $(kt + i\mathbf{i}\mathbf{r} + i\mathbf{j}\tau)$. (It may be thought of as a 3-D 'time' comparable to the 3-D 'mass' represented by $(kE + i\mathbf{i}\mathbf{p} + i\mathbf{j}m)$, and opposed to the 1-D 'charge' and 'space' created by the single well-defined direction for quantized angular momentum.)

The nilpotent wavefunctions demonstrate, among other things, why scalar bosons cannot be massless and why Goldstone bosons are unphysical. They allow an easy calculation of the fine structure terms for the hydrogen atom, of the strong interaction potential (including both the linear potential and the Coulomb term), of the parity states for mesons and baryons, and of propagators for the electromagnetic and other interactions. C, P and T transformations may be performed by a simple operation of the respective quaternion units $\mathbf{j}$, $\mathbf{i}$, $\mathbf{k}$, which shows their connections. The quantum field integrals acting on vacuum are immediately reduced to single nilpotent operators.[18-26] The fermion propagator may be defined without the infrared divergences, while the self-energy of the electron necessarily results in a finite sum.[26]

## 15 The creation of the Dirac state

The Dirac state arises from the fact that the four fundamental parameters, time, space, mass, charge are characterized by the respective algebraic units $i$, **i, j, k**, 1, $i, j, k$. However, the fundamental units of the *algebra* which this entails are created from an anticommuting pentad, with specific rules concerning the two 3-dimensional arrangements, **i, j, k**, $i, j, k$, which are subunits of the general algebra. In reducing the 8 pure units to 5 composite ones, we are constrained by the algebra to map one of them onto the other three *parameters*. So, we may choose to map charge onto space, time and mass, with each of the three charge terms $w$, $s$, $e$, each taking one of the parameters. The composite algebra which emerges is of the form $ki$, $i\mathbf{i}$, $i\mathbf{j}$, $i\mathbf{i}$, $j$, and is isomorphic to the Dirac algebra, though, in creating the Dirac pentads, we have multiplied by an extra $i$ for operational convenience. Diagrammatically, we can represent the process as follows:

| *Time* | *Space* | *Mass* | *Charge* |
|---|---|---|---|
| $i$ | **i  j  k** | 1 | $i\ j\ k$ |
| *Removing charge* | | | |
| $k$ | $i$ | $j$ | |
| *produces* | | | |
| $ik$ | $i\mathbf{i}$  $i\mathbf{j}$  $i\mathbf{k}$ | $j$ | |
| $E$ | **p** | $m$ | |
| *Dirac* | *Dirac* | *Dirac* | |
| *Energy* | *Momentum* | *Rest Mass* | |

Physically, by putting quantized charge components onto time, space and mass units, we introduce quantization to the composite terms, and, since the charges are also conserved quantities, we create a *quantum state* with fixed $E$, **p**, $m$. The concept of 'rest mass' emerges only in this act of 'quantization'. At the same time, the three charges, by being associated with quantities with different mathematical properties (pseudoscalar, vector and real scalar), also take on physical characteristics associated with these quantities, thus breaking the symmetry between them. The symmetry between the weak, strong and electromagnetic interactions is thus broken in the creation of the Dirac state. [27-30]

It is important to stress that the application of quaternion operators in the expression ($ikE + i\mathbf{p} + jm$) or ($kE + ii\mathbf{p} + ijm$) does not in itself create the Dirac state, as the same algebraic expression could be used in a purely mathematical factorization of the classical special relativistic energy-momentum expression. It is the *equating of these operators to the three fundamental charge units*, with their properties of quantization and conservation, that creates the Dirac state by restructuring the meaning of the terms to which they are applied as quantized and conserved ones. This process, as we shall see, implies the direct relation between conservation of charge and conservation of angular momentum which we have predicted must exist purely on the grounds of symmetry.

The Dirac pentad is the largest possible anti-commuting set, and there are exactly six such sets in the algebra. [31] One can always map the gamma algebra to a double-quaternion algebra by defining a triad of anti-commuting numbers (there are 60 such sets) as a quaternion set, then calculating the conjugate triad (another quaternion set) which commutes with the original triad (there are 30 such triad pairs). Such a pair is then the double-quaternion (or quaternion / vector) set, which gives the 3 quaternion, 3 vector, 9 product algebra. It can be shown that, no matter how such as triad pair is chosen, *every* pentad must contain two members of one triad, while the remaining three terms are the product of the third member of the triad with each member of the conjugate triad.

The symmetry breaking occurs when the conjugate triads are chosen. In the general algebra, one is never required to define any triads as privileged, therefore the algebra has 5-dimensional global symmetry. However, once a triad is chosen, the conjugate triad is also uniquely defined. It is this choice which breaks the global symmetry, and allows one to define the $SU(3)$.

## 16 $SU(3)$

The $SU(3)$ symmetry for the strong interaction arises from the vector nature of the term $\mathbf{p}$ in the nilpotent wavefunction. The baryon-type state vector, with three components splitting the three independent components of $\mathbf{p}$, is the only nonzero alternative to the usual fermion nilpotent. The $SU(3)$ symmetry then becomes a straightforward expression of gauge invariance or the perfect symmetry between all the possible phases. The conventional expression for this symmetry requires a covariant derivative of the form:

$$\partial_\mu \rightarrow \partial_\mu + ig_s \frac{\lambda^\alpha}{2} A^{\alpha\mu}(x) \, ,$$

which, in component form, becomes:

$$ip_1 = \partial_1 \to \partial_1 + ig_s \frac{\lambda^\alpha}{2} A^{\alpha 1}(x)$$

$$ip_2 = \partial_2 \to \partial_2 + ig_s \frac{\lambda^\alpha}{2} A^{\alpha 2}(x)$$

$$ip_3 = \partial_3 \to \partial_3 + ig_s \frac{\lambda^\alpha}{2} A^{\alpha 3}(x)$$

$$E = i\partial_0 \to i\partial_0 - g_s \frac{\lambda^\alpha}{2} A^{\alpha 0}(x) \ .$$

Inserting these expressions into the differential form of the baryon state vector, we obtain:

$$\left( \boldsymbol{k} \left( E - g_s \frac{\lambda^\alpha}{2} A^{\alpha 0} \right) \pm \boldsymbol{i} \left( \partial_1 + ig_s \frac{\lambda^\alpha}{2} A^{\alpha 1} \right) + \boldsymbol{ij}\, m \right)$$

$$\left( \boldsymbol{k} \left( E - g_s \frac{\lambda^\alpha}{2} A^{\alpha 0} \right) \pm \boldsymbol{i} \left( \partial_2 + ig_s \frac{\lambda^\alpha}{2} A^{\alpha 2} \right) + \boldsymbol{ij}\, m \right)$$

$$\left( \boldsymbol{k} \left( E - g_s \frac{\lambda^\alpha}{2} A^{\alpha 0} \right) \pm \boldsymbol{i} \left( \partial_3 + ig_s \frac{\lambda^\alpha}{2} A^{\alpha 3} \right) + \boldsymbol{ij}\, m \right).$$

To preserve the nonzero fermionic nilpotent structure, we write this expression in one of the forms:

$$\left( \boldsymbol{k} \left( E - g_s \frac{\lambda^\alpha}{2} A^{\alpha 0} \right) \pm \boldsymbol{i} \left( \partial_1 + ig_s \frac{\lambda^\alpha}{2} \mathbf{A}^\alpha \right) + \boldsymbol{ij}\, m \right) \left( \boldsymbol{k} \left( E - g_s \frac{\lambda^\alpha}{2} A^{\alpha 0} \right) + \boldsymbol{ij}\, m \right) \left( \boldsymbol{k} \left( E - g_s \frac{\lambda^\alpha}{2} A^{\alpha 0} \right) + \boldsymbol{ij}\, m \right)$$

$$\left( \boldsymbol{k} \left( E - g_s \frac{\lambda^\alpha}{2} A^{\alpha 0} \right) + \boldsymbol{ij}\, m \right) \left( \boldsymbol{k} \left( E - g_s \frac{\lambda^\alpha}{2} A^{\alpha 0} \right) \pm \boldsymbol{i} \left( \partial_1 + ig_s \frac{\lambda^\alpha}{2} \mathbf{A}^\alpha \right) + \boldsymbol{ij}\, m \right) \left( \boldsymbol{k} \left( E - g_s \frac{\lambda^\alpha}{2} A^{\alpha 0} \right) + \boldsymbol{ij}\, m \right)$$

$$\left( \boldsymbol{k} \left( E - g_s \frac{\lambda^\alpha}{2} A^{\alpha 0} \right) + \boldsymbol{ij}\, m \right) \left( \boldsymbol{k} \left( E - g_s \frac{\lambda^\alpha}{2} A^{\alpha 0} \right) + \boldsymbol{ij}\, m \right) \left( \boldsymbol{k} \left( E - g_s \frac{\lambda^\alpha}{2} A^{\alpha 0} \right) \pm \boldsymbol{i} \left( \partial_1 + ig_s \frac{\lambda^\alpha}{2} \mathbf{A}^\alpha \right) + \boldsymbol{ij}\, m \right)$$

which are parallel to the six forms incorporated in

$$\psi \sim (BGR - BRG + GRB - GBR + RBG - RGB) \ .$$

The physical interpretation of this representation is that the three quark 'colours' in the baryon are as inseparable as the three dimensions of space and that they stem from exactly the same origin. All 'phases' of the strong interaction are present at the same time, and equally probable, and, if we arbitrarily isolate one phase, the carrier of the 'colour' component of the interaction ($ig_s \lambda^\alpha \mathbf{A}^\alpha / 2$), or the strong charge ($s$), will be 'transferred', at a constant rate, to create the next phase at the same time as the spin (in the $\mathbf{p}$ term); and the current effecting the 'transfer' will be carried by the gluons or generators of the strong field. The 'transfer' of strong charge or 'colour' field, which we describe as the 'strong interaction', is an expression of the innate gauge invariance of the $SU(3)$ representation of the three-part baryon wavefunction, and simultaneously of the conservation of the directional aspect of angular momentum. The interaction, which derives entirely from the nilpotent structure of the baryon wavefunction, will necessarily be nonlocal, and the constant rate of momentum or angular momentum 'transfer', will produce a force which does not depend on the physical separation of the components. Mathematically, this requires a linear potential for the strong force, with an additional Coulomb component needed for spherical symmetry.[24]

## 17 $SU(2)_L \times U(1)$

The conservation properties of the weak and electromagnetic charges appear also to be determined by those of the angular momentum operator, and, in the case of a quark-type arrangement, might be expected to operate the same system of 'privileging' one charge in three during the complete phase cycle (with only one component of angular momentum well-defined). The weak and electric charges, however, are not directly attached to the $\mathbf{p}$ operator, like the strong charge, and so their 'privileged' phases will not necessarily coincide with that of the strong charge or with each other. One of these charges ($w$) is attached to $E$ and the other ($e$) to $m$, and it is the *combination* of these which affects $\mathbf{p}$. For this reason we tend to think of the electric and weak forces as being in some way combined. But these two charges are governed by quite separate symmetries. The charge represented by the quaternion label $\boldsymbol{k}$ (which we call $w$) produces two sign options for $iE$, because the algebra demands complexification of $E$, and there are necessarily two mathematical solutions. Physically, only positive energy exists, because mass-energy is required by the fundamental group symmetry to be a continuum; but the status of the quaternion labels as square roots of $-1$ permits charge conjugation or reversal of the signs of the quaternion labels. So the $-ikE$ states are interpreted as antifermion or charge-conjugated states; and the mass-energy continuum becomes a filled $\boldsymbol{k}$ or weak vacuum for the ground state of the universe, in which such states would not exist.

The filled vacuum is specifically a weak vacuum. It manifests itself through a violation of charge conjugation symmetry for the weak interaction, with consequent violation of either time reversal symmetry or parity. Effectively, this means that, though the weak interaction can tell the difference between particle and antiparticle, it cannot distinguish between + and – signs of weak charge. Where $s$ charges are present, the signs of the $w$ and $s$ charges are linked, though, ideally, they should be independent. A degree of freedom has been removed; and both quarks and free fermions become mixed states, containing +$w$, and suppressed –$w$, states, and involving respective violations of parity and time reversal symmetry. The removal, here, of a degree of freedom from the charges $\pm\ w \pm s \pm e$ coincides with the acquisition of a degree of freedom by $E \pm \mathbf{p}$; in each case the sign of the $\boldsymbol{k}$ operator determines that the Dirac state has the four solutions which result from its quaternionic structure and its 4-D space-time producing the unique $2^{D/2} \times 2^{D/2}$ matrix representation of the Clifford algebra, and, in each case, it requires a filled vacuum relating to the $\boldsymbol{k}$ term.[32]

A violation of parity or time reversal symmetry, consequent upon a violation of charge conjugation, means that only one state of $\boldsymbol{\sigma}.\mathbf{p}$ exists for the pure $w$ interaction for fermions, Because (according to the Dirac equation) $\boldsymbol{\sigma} = -\mathbf{1}$, this is the state of negative helicity or left-handedness. The creation of the alternative states of positive helicity or right-handedness, which are required by the existence of –$\mathbf{p}$ in our formalism, then demands the introduction of mass. In the nilpotent version of the Dirac state, this is associated with the $\boldsymbol{j}$ quaternion label, which defines what we call the electric charge. The presence of $m$ thus simultaneously mixes $E$ and $\mathbf{p}$ terms, right-handed and left-handed components, and the effects of $e$ and $w$ charges.

The existence of separate conservation laws for $w$, $s$, and $e$ charges means that each type of charge must be independent of the other. The characteristic $SU(2)_L$ 'isospin' pattern associated with the weak interaction is an expression of the independence of the weak force from the presence or absence of electric charges. (The electromagnetic $U(1)$ term is, of course, purely a required phase.) The weak interaction must be both uniquely left-handed for fermion states and indifferent to the presence or absence of the electric charge, which introduces the right-handed element. The $SU(2)$ produces a quantum number, $t_3$, such that $(t_3)^2 = (\frac{1}{2})^2$ in half the total number of possible states, that is, in the left-handed ones. For free fermions, the quantum number for the electric force becomes the presence or absence of the electric charge (in the same way as it is the presence or absence of mass). That is, we take 0 and –1 as the quantum numbers ($Q$) (equivalent to the charges 0 and –$e$) for the absence and presence of electric charge (and mass), the – sign being purely historical in origin, and the + sign being reserved for antistates; and so $Q^2 = 1$ in half the total number of possible states (including the right-handed ones). The key electroweak mixing parameter then becomes $\sin^2\theta_W = \Sigma\ (t_3)^2\ /\ \Sigma\ Q^2 = 0.25$.

The weak interaction also has to be indifferent to the presence or absence of the strong charge, that is, to the directional state of the angular momentum operator, and so the same mixing proportion must exist also for quark states, and separately for each 'colour', so that none is preferred. Applying this specifically to quarks, we have the same weak isospin states for one lepton-like 'colour' (that is –1 and 0, or –$e$ and 0), but we also find that the only corresponding isospin states for the other colours that retain both the accepted value of $\sin^2\theta_W$ and the variation of only one 'privileged' quark phase 'instantaneously' in three, are 1 and 0 (or $e$ and 0). In effect, we take the variation 0 0 $-e$ against either an empty background or 'vacuum' (0 0 0) or a full background ($e$ $e$ $e$), so that the two states of weak isospin in the three colours become:

$$
\begin{array}{ccc}
e & e & 0 \\
0 & 0 & -e \;.
\end{array}
$$

With the pattern of $SU(3) \times SU(2)_L \times U(1)$ established from first principles for the strong, weak and electric interactions between fermions, we can, of course, use the already-established formalisms relating to these symmetries to derive Lagrangians, generators, covariant derivatives, and so on, and can extend the reasoning applied to the wavefunctions for $SU(3)$ to those used for $SU(2)_L \times U(1)$. Our immediate concern, here, however, will be to derive charge structures for the fundamental particles.[27-30,33-5]

## 18 The charge structures of quarks and leptons

Fermions (or quarks and leptons) are states characterized by the presence of the weak charge, and the number and structure of all possible fermion states is determined absolutely by the enumeration of all particle states *which are indistinguishable from each other in terms of the weak interaction*. The weak interaction cannot tell, in principle, whether a strong interaction is also operating, and so particles with strong charges (quarks) ought to be indistinguishable by this interaction from particles without strong charges (leptons). In the case of quarks, it cannot tell the difference between a filled 'electromagnetic vacuum' (weak isospin up state) and an empty one (weak isospin down state). The weak interaction, in addition, is also indifferent to the sign of the weak charge, and responds (via the vacuum) only to the status of fermion or antifermion; this produces mixing between the respective fermion *generations*, defined with $+w$, with $-w$ and P violation, and with $-w$ and T violation. From these sets of equally probable states (excluding energy considerations), we define all the possible distinctions between fermion / antifermion; quark / lepton; isospin up / isospin down; and the three quark-lepton generations. The distinctions are made in terms of the other charges ($s$ and $e$), and of mass.

A representation of this process can be made in terms of conservation of angular momentum, which we have already associated with the conservation of each of the charges. If we take $\boldsymbol{\sigma}.\hat{\mathbf{p}}$ (or $-\boldsymbol{\sigma}.\hat{\mathbf{p}}$, using the historically-established sign conventions for charges) as equivalent in unit charge terms to an expression in which $\hat{\mathbf{p}}$ becomes the unit vector components $\hat{\mathbf{p}}_1$, $\hat{\mathbf{p}}_2$, $\hat{\mathbf{p}}_3$, in successive phases of the strong interaction, and apply this to the strong charge quaternion operator $\boldsymbol{i}$, the units of strong charge will become $0\boldsymbol{i}$ or $1\boldsymbol{i}$, depending on the supposed instantaneous direction of the angular momentum vector. Only one component of a baryon will have this unit at any instant. In reality, of course, gauge invariance ensures that all possible phases exist at once, so spin becomes a property of the entire system and not of the component quarks.

The same angular momentum term ($\boldsymbol{\sigma}.\hat{\mathbf{p}}$) carries the information concerning the conservation of the other two charge terms, and, in effect, the three charges are separately conserved because they represent three different aspects of the angular momentum conservation process. In the case of the weak and electric charges, the random unit vector components $\hat{\mathbf{p}}_1$, $\hat{\mathbf{p}}_2$, $\hat{\mathbf{p}}_3$, are associated respectively with the sign, and the magnitude of the angular momentum state, through the connections of $\mathbf{p}$ with $E$ and $\mathbf{p}$ with $m$. We can, in fact, generalise the procedure by applying $\boldsymbol{\sigma}.\hat{\mathbf{p}}_1$, $\boldsymbol{\sigma}.\hat{\mathbf{p}}_2$, $\boldsymbol{\sigma}.\hat{\mathbf{p}}_3$ to the quaternion operators ($\boldsymbol{k}$ and $\boldsymbol{j}$) specifying $w$ and $e$, but with the sequence of unit vectors determined separately in each case. It is then the various alignments between the sequences of unit vectors or *phases* applied to $s$, $w$ and $e$ which determine the nature of the fermion state produced.

For example, if we align the unit vectors applied to $w$ and $e$, we are effectively aligning the $E$ and $m$ phases with each other, and so necessarily with the $\mathbf{p}$ phase, which means that the system has a single phase and so cannot be baryonic. The $\mathbf{p}$ phase is defined with $E$ and $m$, and there is no strong charge. We have thus defined a free fermion or lepton. In a baryon system, with strong charges present, the vectors assigned to the weak and electric charges, and hence to $E$ and $m$, will not be aligned, and, consequently, the $\mathbf{p}$ phase is not fixed with respect to them.

To complete the representation of all possible fermions, we need to incorporate the effect of weak isospin and the parity- and time-reversal-violation of the second and third generations. Reversal of isospin is accomplished by replacing a term such as $-\boldsymbol{j}\hat{\mathbf{p}}_1$ with $-\boldsymbol{j}(\hat{\mathbf{p}}_1 - \mathbf{1})$, the $\boldsymbol{j}\mathbf{1}$ representing the filled 'electric vacuum' state. Charge conjugation violation may be represented by the non-algebraic symbols $z_P$ and $z_T$, depending on whether it is accompanied by $P$ or $T$ violation. In using these symbols, we are merely saying that we are treating the $-w$ of the second and third generator as though it were positive in the same way as the $w$ of the first generation. We can now express quark structures in the following form:

| | |
|---|---|
| down | $-\boldsymbol{\sigma}.\,(-\boldsymbol{j}\hat{\mathbf{p}}_a + \boldsymbol{i}\hat{\mathbf{p}}_b + \boldsymbol{k}\hat{\mathbf{p}}_c)$ |
| up | $-\boldsymbol{\sigma}.\,(-\boldsymbol{j}(\hat{\mathbf{p}}_a - \mathbf{1}) + \boldsymbol{i}\hat{\mathbf{p}}_b + \boldsymbol{k}\hat{\mathbf{p}}_c)$ |
| strange | $-\boldsymbol{\sigma}.\,(-\boldsymbol{j}\hat{\mathbf{p}}_a + \boldsymbol{i}\hat{\mathbf{p}}_b - z_P\boldsymbol{k}\hat{\mathbf{p}}_c)$ |
| charmed | $-\boldsymbol{\sigma}.\,(-\boldsymbol{j}(\hat{\mathbf{p}}_a - \mathbf{1}) + \boldsymbol{i}\hat{\mathbf{p}}_b - z_P\boldsymbol{k}\hat{\mathbf{p}}_c)$ |
| bottom | $-\boldsymbol{\sigma}.\,(-\boldsymbol{j}\hat{\mathbf{p}}_a + \boldsymbol{i}\hat{\mathbf{p}}_b - z_T\boldsymbol{k}\hat{\mathbf{p}}_c)$ |
| top | $-\boldsymbol{\sigma}.\,(-\boldsymbol{j}(\hat{\mathbf{p}}_a - \mathbf{1}) + \boldsymbol{i}\hat{\mathbf{p}}_b - z_T\boldsymbol{k}\hat{\mathbf{p}}_c)$ |

Here, $-j$ represents electric charge (conventionally negative), $i$ strong, $k$ weak. $a$, $b$, $c$ are *each* randomly 1, 2, 3, except that $b \neq c$. Both $- z_P k$ and $- z_T k$ become equivalent to $k$, for the purposes of the weak interaction. For the corresponding leptons, the components are all in phase ($\hat{\mathbf{p}}_a$), and there is no directional component:

| | |
|---|---|
| electron | $-\boldsymbol{\sigma} . (-j\hat{\mathbf{p}}_a + k\hat{\mathbf{p}}_a)$ |
| $e$ neutrino | $-\boldsymbol{\sigma} . (-j(\hat{\mathbf{p}}_a - \mathbf{1}) + k\hat{\mathbf{p}}_a)$ |
| muon | $-\boldsymbol{\sigma} . (-j\hat{\mathbf{p}}_a - z_P k\hat{\mathbf{p}}_a)$ |
| $\mu$ neutrino | $-\boldsymbol{\sigma} . (-j(\hat{\mathbf{p}}_a - \mathbf{1}) - z_P k\hat{\mathbf{p}}_a)$ |
| tau | $-\boldsymbol{\sigma} . (-j\hat{\mathbf{p}}_a - z_T k\hat{\mathbf{p}}_a)$ |
| $\tau$ neutrino | $-\boldsymbol{\sigma} . (-j(\hat{\mathbf{p}}_a - \mathbf{1}) - z_T k\hat{\mathbf{p}}_a)$ |

Both antiquarks and antileptons simply replace $-\boldsymbol{\sigma}$ with $\boldsymbol{\sigma}$.

## 19 Quarks and leptons

From these formulae, the 0 and 1 charge structures of the fundamental fermions may be expressed in terms of a set of three 'quark' tables, A-C, with an extra table L for the left-handed leptons and antileptons:

**A**

| | | B | G | R |
|---|---|---|---|---|
| u | $+e$ | 1$j$ | 1$j$ | 0$i$ |
| | $+s$ | 1$i$ | 0$k$ | 0$j$ |
| | $+w$ | 1$k$ | 0$i$ | 0$k$ |
| | | | | |
| d | $-e$ | 0$j$ | 0$k$ | 1$j$ |
| | $+s$ | 1$i$ | 0$i$ | 0$k$ |
| | $+w$ | 1$k$ | 0$j$ | 0$i$ |
| | | | | |
| c | $+e$ | 1$j$ | 1$j$ | 0$i$ |
| | $+s$ | 1$i$ | 0$k$ | 0$j$ |
| | $-w$ | $z_P\boldsymbol{k}$ | 0$i$ | 0$k$ |
| | | | | |
| s | $-e$ | 0$j$ | 0$k$ | 1$j$ |
| | $+s$ | 1$i$ | 0$i$ | 0$k$ |
| | $-w$ | $z_P\boldsymbol{k}$ | 0$j$ | 0$i$ |
| | | | | |
| t | $+e$ | 1$j$ | 1$j$ | 0$i$ |
| | $+s$ | 1$i$ | 0$k$ | 0$j$ |
| | $-w$ | $Z_T\boldsymbol{k}$ | 0$i$ | 0$k$ |
| | | | | |
| b | $-e$ | 0$j$ | 0$k$ | 1$j$ |
| | $+s$ | 1$i$ | 0$i$ | 0$k$ |
| | $-w$ | $Z_T\boldsymbol{k}$ | 0$j$ | 0$i$ |
| | | | | |

**B**

| | | B | G | R |
|---|---|---|---|---|
| u | $+e$ | 1$j$ | 1$j$ | 0$k$ |
| | $+s$ | 0$i$ | 0$k$ | 1$i$ |
| | $+w$ | 1$k$ | 0$i$ | 0$j$ |
| | | | | |
| d | $-e$ | 0$i$ | 0$k$ | 1$j$ |
| | $+s$ | 0$j$ | 0$i$ | 1$i$ |
| | $+w$ | 1$k$ | 0$j$ | 0$k$ |
| | | | | |
| c | $+e$ | 1$j$ | 1$j$ | 0$k$ |
| | $+s$ | 0$i$ | 0$k$ | 1$i$ |
| | $-w$ | $z_P\boldsymbol{k}$ | 0$i$ | 0$j$ |
| | | | | |
| s | $-e$ | 0$i$ | 0$k$ | 1$j$ |
| | $+s$ | 0$j$ | 0$i$ | 1$i$ |
| | $-w$ | $z_P\boldsymbol{k}$ | 0$j$ | 0$k$ |
| | | | | |
| t | $+e$ | 1$j$ | 1$j$ | 0$k$ |
| | $+s$ | 0$i$ | 0$k$ | 1$i$ |
| | $-w$ | $Z_T\boldsymbol{k}$ | 0$i$ | 0$j$ |
| | | | | |
| b | $-e$ | 0$i$ | 0$k$ | 1$j$ |
| | $+s$ | 0$j$ | 0$i$ | 1$i$ |
| | $-w$ | $Z_T\boldsymbol{k}$ | 0$j$ | 0$k$ |
| | | | | |

| | | B | G | R |
|---|---|---|---|---|
| u | $+e$ | 1$j$ | 1$j$ | 0$k$ |
| | $+s$ | 0$i$ | 1$i$ | 0$j$ |
| | $+w$ | 1$k$ | 0$k$ | 0$i$ |
| | | | | |
| d | $-e$ | 0$j$ | 0$k$ | 1$j$ |
| | $+s$ | 0$i$ | 1$i$ | 0$k$ |
| | $+w$ | 1$k$ | 0$j$ | 0$i$ |
| | | | | |
| c | $+e$ | 1$j$ | 1$j$ | 0$k$ |
| | $+s$ | 0$i$ | 1$i$ | 0$j$ |
| | $-w$ | $z_P\textbf{\textit{k}}$ | 0$k$ | 0$i$ |
| | | | | |
| s | $-e$ | 0$j$ | 0$k$ | 1$j$ |
| | $+s$ | 0$i$ | 1$i$ | 0$k$ |
| | $-w$ | $z_P\textbf{\textit{k}}$ | 0$j$ | 0$i$ |
| | | | | |
| t | $+e$ | 1$j$ | 1$j$ | 0$k$ |
| | $+s$ | 0$i$ | 1$i$ | 0$j$ |
| | $-w$ | $Z_T\textbf{\textit{k}}$ | 0$k$ | 0$i$ |
| | | | | |
| b | $-e$ | 0$j$ | 0$k$ | 1$j$ |
| | $+s$ | 0$i$ | 1$i$ | 0$k$ |
| | $-w$ | $Z_T\textbf{\textit{k}}$ | 0$j$ | 0$i$ |
| | | | | |

| | | $\overline{e}$ | $\overline{e}$ | $\nu_e$ |
|---|---|---|---|---|
| | $+e$ | 1$j$ | 1$j$ | 0$j$ |
| | $+s$ | 0$k$ | 0$i$ | 0$i$ |
| | $+w$ | 0$i$ | 0$k$ | 1$k$ |
| | | | | $e$ |
| | $-e$ | 0$i$ | 0$k$ | 1$j$ |
| | $+s$ | 0$j$ | 0$i$ | 0$i$ |
| | $+w$ | 0$k$ | 0$j$ | 1$k$ |
| | | $\overline{\mu}$ | $\overline{\mu}$ | $\nu_u$ |
| | $+e$ | 1$j$ | 1$j$ | 0$j$ |
| | $+s$ | 0$k$ | 0$i$ | 0$i$ |
| | $-w$ | 0$i$ | 0$k$ | $z_P\textbf{\textit{k}}$ |
| | | | | $\mu$ |
| | $-e$ | 0$i$ | 0$k$ | 1$j$ |
| | $+s$ | 0$j$ | 0$i$ | 0$i$ |
| | $-w$ | 0$k$ | 0$j$ | $z_P\textbf{\textit{k}}$ |
| | | $\overline{\tau}$ | $\overline{\tau}$ | $\nu_\tau$ |
| | $+e$ | 1$j$ | 1$j$ | 0$j$ |
| | $+s$ | 0$k$ | 0$i$ | 0$i$ |
| | $-w$ | 0$i$ | 0$k$ | $Z_T\textbf{\textit{k}}$ |
| | | | | $\tau$ |
| | $-e$ | 0$i$ | 0$k$ | 1$j$ |
| | $+s$ | 0$j$ | 0$i$ | 0$i$ |
| | $-w$ | 0$k$ | 0$j$ | $Z_T\textbf{\textit{k}}$ |

Applying these to the known fermions, A-C would appear to represent the coloured quark system, with $s$ pictured as being 'exchanged' between the three states (although in reality, of course, all the states exist simultaneously), in the same way as the operator **p** in the nilpotent baryon wavefunction. In relation to these tables, we can look on symmetry-breaking, in general, as a consequence of the setting up of the algebraic model for charges. When we map time, space and mass onto the charges $w$-$s$-$e$, to create the anticommuting Dirac pentad, only one charge ($s$) has the full range of vector options. 'Fixing' one of the others (say $e$) for $s$ to vary against, gives us only 2 remaining options for $w$, unit on the same colour as $e$ or unit on a different one. Putting both $w$ and $e$ on the same colour denies the necessary three degrees of freedom in the direction of angular momentum, so this is forbidden in a quark system. In principle, there are also two other quark tables (D-E), but, with the application of the exclusion rules defined above, these both reduce to the lepton table L. The reduction from A-E to A-C plus L can be thought of as similar to a reduction from the full Dirac pentad to a 4-vector or quaternion representation.

|   |   | **D** |   |   |
| --- | --- | --- | --- | --- |
|   |   | **B** | **G** | **R** |
| u | $+e$ | 1$j$ | 1$j$ | 0$i$ |
|   | $+s$ | 0$k$ | 1$i$ | 0$j$ |
|   | $+w$ | 0$i$ | 0$k$ | 1$k$ |
|   |   |   |   |   |
| d | $-e$ | 0$i$ | 0$k$ | 1$j$ |
|   | $+s$ | 0$j$ | 1$i$ | 0$i$ |
|   | $+w$ | 0$k$ | 0$j$ | 1$k$ |
|   |   |   |   |   |
| c | $+e$ | 1$j$ | 1$j$ | 0$i$ |
|   | $+s$ | 0$k$ | 1$i$ | 0$j$ |
|   | $-w$ | 0$i$ | 0$k$ | $z_P\mathbf{k}$ |
|   |   |   |   |   |
| s | $-e$ | 0$i$ | 0$k$ | 1$j$ |
|   | $+s$ | 0$j$ | 1$i$ | 0$i$ |
|   | $-w$ | 0$k$ | 0$j$ | $z_P\mathbf{k}$ |
|   |   |   |   |   |
| t | $+e$ | 1$j$ | 1$j$ | 0$i$ |
|   | $+s$ | 0$k$ | 1$i$ | 0$j$ |
|   | $-w$ | 0$i$ | 0$k$ | $Z_T\mathbf{k}$ |
|   |   |   |   |   |
| b | $-e$ | 0$i$ | 0$k$ | 1$j$ |
|   | $+s$ | 0$j$ | 1$i$ | 0$i$ |
|   | $-w$ | 0$k$ | 0$j$ | $Z_T\mathbf{k}$ |
|   |   |   |   |   |

|   |   | **E** |   |   |
| --- | --- | --- | --- | --- |
|   |   | **B** | **G** | **R** |
| u | $+e$ | 1$j$ | 1$j$ | 0$j$ |
|   | $+s$ | 0$k$ | 0$i$ | 1$i$ |
|   | $+w$ | 0$i$ | 0$k$ | 1$k$ |
|   |   |   |   |   |
| d | $-e$ | 0$i$ | 0$k$ | 1$j$ |
|   | $+s$ | 0$j$ | 0$i$ | 1$i$ |
|   | $+w$ | 0$k$ | 0$j$ | 1$k$ |
|   |   |   |   |   |
| c | $+e$ | 1$j$ | 1$j$ | 0$j$ |
|   | $+s$ | 0$k$ | 0$i$ | 1$i$ |
|   | $-w$ | 0$i$ | 0$k$ | $z_P\mathbf{k}$ |
|   |   |   |   |   |
| s | $-e$ | 0$i$ | 0$k$ | 1$j$ |
|   | $+s$ | 0$j$ | 0$i$ | 1$i$ |
|   | $-w$ | 0$k$ | 0$j$ | $z_P\mathbf{k}$ |
|   |   |   |   |   |
| t | $+e$ | 1$j$ | 1$j$ | 0$j$ |
|   | $+s$ | 0$k$ | 0$i$ | 1$i$ |
|   | $-w$ | 0$i$ | 0$k$ | $Z_T\mathbf{k}$ |
|   |   |   |   |   |
| b | $-e$ | 0$i$ | 0$k$ | 1$j$ |
|   | $+s$ | 0$j$ | 0$i$ | 1$i$ |
|   | $-w$ | 0$k$ | 0$j$ | $Z_T\mathbf{k}$ |
|   |   |   |   |   |

An important aspect of this model is the fact that the charges ($e$, $s$, w) are irrotational, but the quaternion labels ($j$, $i$, $k$) are not, and this has been used in previous work as the primary derivation of the tables. Even here, E appears to be excluded automatically by requiring all three quaternions to be attached to specified charges, thus losing the three required degrees of rotational freedom, and, at the same time, necessarily violating Pauli exclusion. Significantly, also, though the separate and random variation of $\hat{\mathbf{p}}_a$, $\hat{\mathbf{p}}_b$, and $\hat{\mathbf{p}}_c$, in a quark-based system suggests 27 possible variations in charge structure, the number of different outcomes is reduced by repetitions, and is five, as in the anticommuting pentads of the Dirac algebra. Effectively, we 'privilege' one of $\hat{\mathbf{p}}_a$, $\hat{\mathbf{p}}_b$, $\hat{\mathbf{p}}_c$, by allowing it complete variation with respect to the others; $\hat{\mathbf{p}}_b$ is the one selected, and it corresponds to the 'privileging' of $\mathbf{p}$ as a vector term with full variation in the Dirac anticommuting pentad.

In its most compact form, the entire set of charge structures for fermions and antifermions may be derived from a single expression:

$$\boldsymbol{\sigma}_z.(\boldsymbol{i}\, \hat{\mathbf{p}}_a\, (\delta_{bc} - 1) + \boldsymbol{j}\, (\hat{\mathbf{p}}_b - \mathbf{1}\delta_{0m}) + \boldsymbol{k}\, \hat{\mathbf{p}}_c\, (-1)^{\delta}1g\, g)$$

Here, $\boldsymbol{\sigma}_z$ is the spin pseudovector component defined in the $z$ direction; and the units of quantized angular momentum, $\hat{\mathbf{p}}_a$, $\hat{\mathbf{p}}_b$, $\hat{\mathbf{p}}_c$ are selected *randomly* and *independently* from the three orthogonal components $\hat{\mathbf{p}}_x$, $\hat{\mathbf{p}}_y$, $\hat{\mathbf{p}}_z$. Each of the other operators creates one of the fundamental divisions in fermionic structure – fermion / antifermion; quark / lepton (colour); weak up isospin / weak down isospin; and the three generations – which are identified, respectively, by the weak, strong, electromagnetic and gravitational interactions. $\boldsymbol{\sigma}_z = -\mathbf{1}$ defines left-handed states; $\boldsymbol{\sigma}_z = \mathbf{1}$ defines right-handed. Assuming a filled weak vacuum, left-handed states become predominantly fermionic, while right-handed states are antifermionic 'holes' in the vacuum. $b = c$ produces leptons; $b \neq c$ produces quarks. Taking into account all three directions at once, when $b \neq c$, we define baryons composed of three quarks (and mesons composed of quark and antiquark), in which each of $a$, $b$, $c$ cycle through the directions $x, y, z$.

$m$ is an electromagnetic mass unit, which selects the state of weak isospin. It becomes 1 when the mass is present and 0 it is when absent. The unit condition can be taken as an empty electromagnetic vacuum; the zero condition a filled one. $g$ represents a conjugation of weak charge units, with $g = -1$ meaning maximal conjugation. If conjugation fails maximally, then $g = 1$. $g$ can also be taken as a composite term, containing a parity element ($P$) and a time-reversal element ($T$), providing two ways for the conjugated $PT$ to remain at the unconjugated value (1). $g = -1$ produces the generation $u, d, \nu_e, e$; $g = 1$, with $P$ violated, produces $c, s, \nu_\mu, \mu$; $g = 1$, with $T$ violated, produces $t, b, \nu_\tau, \tau$.

The charge conjugation from $-w$ to $w$, in the second and third generations, which is represented in the tables by $z_P$ or $z_T$, is brought about by the filled weak vacuum needed to avoid negative energy states. The two weak isospin states are associated with this idea, the $\mathbf{1}$ in $(\hat{\mathbf{p}}_b - \mathbf{1}\delta_{0m})$ being a 'filled' state, with its absence an unfilled state, and the weak interaction acts by annihilating and creating $e$, either filling the vacuum or emptying it – which is why, unlike the strong interaction, it always involves the equivalent of particle + antiparticle = particle + antiparticle, and involves a massive intermediate boson. We thus create two possible vacuum states to allow variation of the sign of electric charge by weak isospin, and this variation is linked to the filling of the vacuum which occurs in the weak interaction. The weak and electric interactions are linked by this filled vacuum in the $SU(2)_L \times U(1)$ model, as they are in our description of weak isospin, and we can regard these as alternative formalisms for representing the same physical truth. It is significant that the Higgs mechanism for generating masses of intermediate weak bosons and fermions requires the same Higgs vacuum field both for $SU(2)_L$ and for $U(1)$.

## 20 Conservation of type of charge and conservation of angular momentum

In generating the particle charge structures, we have established the connection between the conservation of the type of charge and the conservation of angular momentum, which emerges from the fundamental symmetry of space, time, mass and charge. Each type of charge (strong, weak and electromagnetic) is related to a separate aspect of angular momentum conservation (direction, orientation (sign), and magnitude). The angular momentum operator in the nilpotent is also vital to the meaning of quantum mechanics itself, as it effectively defines the classical / quantum transition. Real physical processes involve discrete energy transfer between discrete charged or massive particles. In quantum mechanical terms, discrete energy transfer involves collapse of the wavefunction. This collapse may be considered as a change which breaks the direct connection between the conservation laws of type of charge and angular momentum. Where the connection is maintained in a coherent way, even on a large scale, we have quantum mechanics. Where it is not, we have decoherence through the vector addition of the noncoherent individual $\mathbf{p}$ components, and, therefore, energy transfer. (If the $\mathbf{p}$ terms remain aligned, there is no need to alter $E$.) The same also applies for a change of charge structure or rest mass state. The level of decoherence is measured by the increase in entropy (or the number of noncoherent states), and any process involving an interaction between two fermions will involve some measure of it. In practical terms, making a classical measurement means the application of fields of sufficient size to make the whole system decohere and reduce any quantum mechanical variation in spatial coordinates during a given time interval to the level of the uncertainty principle.

## 21 The lepton-like quark model

It seems evident, from the algebraic representation, that the most appropriate mathematical model to represent quarks is an extension of the integrally-charged coloured quark model, originally proposed by Han and Nambu in 1965, and operating to produce observed fractional charges in much the same way as the parallel phenomenon of the fractional quantum Hall effect in condensed matter physics.[36] In a *fully* gauge invariant theory of the strong interaction, with quark colours intrinsically inseparable, the underlying charges could easily be integral while always being perceived as fractional in effect at any energy. With *perfect* gauge invariance between the coloured phases, and hence perfect infrared slavery (as our model requires), there will be no transition regime between implicit and explicit colour, and no finite energy range at which integral charges or colour properties will emerge. Observed charges, measured by quantum electrodynamics, will be exactly fractional at all energies.

The advantages of this version of quark structure are that it is already lepton-like, allowing for an easier unification of the two types of fermion; that it makes $\sin^2\theta_W = 0.25$ for both leptons *and* quarks (which is close to the observed value); that it predicts an exact Grand Unification of all four forces at the Planck mass; that it avoids difficulties of the definition of a charge unit if both 2/3 and 1/3 are fundamental units; and that it produces *two* values for weak hypercharge, as required, when the Higgs mechanism is applied to the creation of quark masses, and they are also the correct values (±1).[30,35,37] A model based on real fractional charges does none of these, and it fails also to explain how a fermion can 'know' whether it is a quark or a lepton by any means other than the strong interaction.

The model also explains easily how mass is generated when an element of partial right-handedness is introduced into an intrinsically left-handed system. In principle, anything which alters the signs of the terms in the expression $(i\ \hat{\mathbf{p}}_a\ (\delta_{bc} - 1) + \boldsymbol{j}\ (\hat{\mathbf{p}}_b - \mathbf{1}\delta_{bm}) + \boldsymbol{k}\ \hat{\mathbf{p}}_c\ (-1)^{\delta}1g\ g)$, or reduces any of these terms to zero, is a mass generator, because it is equivalent to introducing the opposite sign of $\boldsymbol{\sigma}_z$ or a partially right-handed state. Thus mass can be produced separately by weak isospin, by quark confinement, and by weak charge conjugation violation. Various calculations have been made, using these mechanisms, of the masses of baryons and mesons, quarks and leptons.[2,4,27-8,30]

Using the tables A-C and L, we can explain many facts related to particle physics, and also make some new predictions. For example, if we derive baryon and meson charge structures from those of the component quarks and antiquarks, we find that all baryons have a weak charge structure of *w*, while all mesons have a weak charge structure 0, with the exception of states like the *K* mesons, and the other meson states combining a fermion and antifermion of two different generations. For these states, we find alternative weak charge structures of 0 or $\pm (1 + z)w$, where $z$ may be $z_P$ or $z_T$. (Mesons combining the second and third generations will have structure combining $z_P$ and $z_T$.) The alternatives depend on the particular colour-anticolour combination, and particular quark representation, we choose. Clearly, 0 and $\pm (1 + z)w$ have to be indistinguishable in weak terms, but we have already broken a symmetry (either P or T) in creating $z$, which means that we are also obliged to break another to maintain overall CPT invariance. In the case of a bosonic state, charge conjugation must be preserved, so we are obliged to break CP (or T) as well as P, or CT (or P) as well as T. Our prediction is that such an additional violation will be found in all states of this kind.[2,4] It has already been observed in $K^o$ and $\overline{K}^o$ mixed states, and is now known to occur in $K^o$ and $\overline{K}^o$, taken separately, as well as being extended to incorporate mesons combining first and third generation components. However, it should also be observable in $K^+$ and $K^-$, and in the equivalent states in other generations. It should also be observed in the weak decays of Bose-Einstein condensates, which again involve a $(1 + z)w$, weak charge structure.

Some other predictions may be made of a more quantitative nature. For example, from $\sin^2\theta_W = 0.25$ at the energy of the $Z^0$ boson (which theory suggests will measure slightly lower if it involves the emergence of the boson itself), we derive $\sin^2\theta_W = 1$ at Grand Unification, and a Grand Unification energy equivalent to the Planck mass. This leads to a testable prediction for $1/\alpha$ at 118 at 14 TeV, with appropriate variation up to that value.[28,34,36] (This is measurably lower than the prediction of 125 at 14 TeV from minimal $SU(5)$.) There are grounds (though much more tentative) for suggesting that the mass of the Higgs boson is of order 182 GeV or $2M_Z$, which suggests that the most favoured decay will be via $2Z^0$ or the four lepton route.[28,30,32-4] A separate prediction of an accelerating cosmological redshift, with a deceleration parameter –1, follows from the gravitation-inertia interpretation of general relativity.[12-13,23]

## 22 Grand unification

The idea that the 5-fold Dirac algebra is responsible for the symmetry breaking which leads to the $SU(3) \times SU(2)_L \times U(1)$ splitting in the interactions between fundamental particles, suggests that Grand Unification may involve the $SU(5)$ group, or even $U(5)$. In principle, we derive five representations of the electric, strong and weak charge states (A-E), which map onto the charge units ($e$, $s$, $w$), and the five quantities ($m$, **p**, $E$) involved in the Dirac equation.[20-1,28-30,37] The 24 $SU(5)$ generators can be represented in terms of any of these units, for example:

|       | $\bar{s_G}$ | $\bar{s_B}$ | $\bar{s_R}$ | $\bar{w}$ | $\bar{e}$ |
|-------|------|------|------|------|------|
| $s_G$ |      |      |      |      |      |
| $s_B$ |      | 8 Gluons |  | $Y$ | $X$ |
| $s_R$ |      |      |      |      |      |
| $w$   |      | $Y$  |      | $Z^0, \gamma$ | $W^-$ |
| $e$   |      | $X$  |      | $W^+$ | $Z^0, \gamma$ |

and

|       | $\bar{p_x}$ | $\bar{p_y}$ | $\bar{p_z}$ | $\bar{E}$ | $\bar{m}$ |
|-------|------|------|------|------|------|
| $p_x$ |      |      |      |      |      |
| $p_y$ |      | 8 Gluons |  | $Y$ | $X$ |
| $p_z$ |      |      |      |      |      |
| $E$   |      | $Y$  |      | $Z^0, \gamma$ | $W^-$ |
| $m$   |      | $X$  |      | $W^+$ | $Z^0, \gamma$ |

With an additional 25$^{th}$ generator, linking all the colourless and charge-free bosons along the diagonal, the group would become $U(5)$. This generator, if it existed, would couple to all matter in proportion to the amount, and, as a colour singlet, would be ubiquitous. This is a precise description of the force of gravity, and, since we have shown that Grand Unification of the electromagnetic, strong and weak forces occurs at the Planck mass, the energy usually taken as characteristic of quantum gravity, then it is highly probable that $U(5)$ group is the true Grand Unification group, and that it also incorporates a gravity generator – though most probably a spin 1 generator for the inertial reaction, rather than a spin 2, or other hypothetical, generator for gravity itself.

The Planck mass is also the likely candidate for the cut-off energy which ensures the finite summation of fermion self-energies required by the nilpotent formulation. Through the need for a filled vacuum and the continuous nature of mass-energy, gravity is, we believe, the instantaneous carrier of the wavefunction correlations involved in nonlocality, and the Planck mass is the quantum of the inertial interactions, which we have elsewhere found to be the result of the effect of gravity on the time-delayed nature of nongravitational interactions.[13-16] These in turn produce the inertial masses associated with charged particles, by a coupling to the Higgs field which fills the vacuum state.[26]

A $U(5)$ Grand Unification would have the advantage of making all the generators become pure phases, and identical in form, at the Grand Unification energy. The most likely possibility (especially if it concerns the inertial reaction) is that the Grand Unification energy would become a kind of 'event horizon', or unattainable limit, at which $\sin^2\theta_W$ would equal 1, and separate conservation laws for charges would have no meaning. In fact, the necessity for separate conservation would prohibit its attainment, as it already prohibits direct proton decay. (The $X$ and $Y$ generators do not, we believe, imply direct proton decay, as they are linked to the particle + antiparticle mechanism of the ordinary weak interaction.)

The five component terms of the Dirac pentad ($E$-**p**-$m$) are discrete, having become so in the process of superimposing the charges $w$, $s$, $e$ on the original group parameters time, space and mass. The discreteness may be thought of, in some sense, as being due to complexification. A nilpotent Dirac pentad is necessarily complex, and complexity is also necessary to fermions because the complex term $\mathbf{i}kE$ (occupying the same position in the formalism as $w$) is intrinsic to their definition. Complexity creates a discontinuous $U(5)$ because in the $E$-**p**-$m$ pentad we have a real quantized mass and an imaginary quantized $E$. The group is also $U(5)$ because the gravity operator ('mass') has been automatically included in the process of 'compactification', or reduction of the eight original units ($1, \mathbf{i}, \mathbf{j}, \mathbf{k}, i, \mathbf{i}, \mathbf{j}, \mathbf{k}$) to the five composite Dirac terms ($\mathbf{i}k, \mathbf{i}i, \mathbf{i}j, \mathbf{i}k, \mathbf{j}$). Complex $i$ introduces discreteness in the square-rooting process which produces $iE$, and this links up with the fermion / boson distinction.

### 23 The Dirac algebra operators and *SU*(5) generators

Essentially, there are nine elements of the Dirac algebra which contain a vector and a quaternion term (and another nine if it is made complex). However, the 32-part algebra requires only five primitives, only three of which are vector quaternions. In principle, the system inevitably privileges one set of three out of the nine – only one quaternion is allowed to have a vector in the primitive set (leading to the behaviour of **p** and *s*). Where the system is complex, there seems also to be a requirement for some of the primitive terms to have a coefficient of imaginary *i* and others to have a real coefficient. We can draw up the following multiplication table for the components of the algebra:

| | | *ii* | *ij* | *ik* | *j* | *ik* |
|---|---|---|---|---|---|---|
| | | | | | | |
| *ii* | | –1 | – *i***k** | *i***j** | *k***i** | – *ij***i** |
| *ij* | | *i***k** | –1 | – *i***i** | *k***j** | – *ij***j** |
| *ik* | | – *i***j** | *i***i** | –1 | *k***k** | – *ij***k** |
| *j* | | – *k***i** | – *k***j** | – *k***k** | –1 | *i***i** |
| *ik* | | *ij***i** | *ij***j** | *ij***k** | – *i***i** | 1 |
| | | | | | | |

By comparison with the tables for the *SU*(5) generators, *ii*, *ij*, *ik* represents the strong charge in its three-colour form, *j* the electric charge, and *ik* the weak charge. The terms ± *ii*, ± *i***j**, ± *i***k** represent the colour non-singlet gluons; ± *ii* the weak generators *W*⁺ and *W*⁻; and ± *k***i**, ± *k***j**, ± *k***k** and ± *ij***i**, ± *ij***j**, ± *ij***k** the generators *X* and *Y* of the Grand Unified theory. All possible parts of the Dirac algebra are represented as either source terms or generators, if we take no account of the complex factor *i*. (It is notable that terms appear as sources or generators only with this factor or without it – there is no duplication, and the total number of possible outcomes remains 16.) The five zero-charge generators along the diagonal, the two colourless gluons, Z⁰, γ, and the gravity operator may be assumed completely mixed at Grand Unification in an overall *U*(5) structure.

This is what happens if we assume the charge accommodation produces a complete one-to-one correspondence with the Dirac algebra. If we now choose to construct a system for charge accommodation based on a double quaternion algebra (i.e. replacing the vector set **i**, **j**, **k** with a second, independent, quaternion set, using the same symbols), and, leaving out the *i* term entirely, we obtain the following:

| | | $i\mathbf{i}$ | $i\mathbf{j}$ | $i\mathbf{k}$ | $j$ | $k$ |
|---|---|---|---|---|---|---|
| | | | | | | |
| $i\mathbf{i}$ | | 1 | $-\mathbf{k}$ | $\mathbf{j}$ | $k\mathbf{i}$ | $-j\mathbf{i}$ |
| $i\mathbf{j}$ | | $\mathbf{k}$ | 1 | $-\mathbf{i}$ | $k\mathbf{j}$ | $-j\mathbf{j}$ |
| $i\mathbf{k}$ | | $-\mathbf{j}$ | $\mathbf{i}$ | 1 | $k\mathbf{k}$ | $-j\mathbf{k}$ |
| $j$ | | $-k\mathbf{i}$ | $-k\mathbf{j}$ | $-k\mathbf{k}$ | $-1$ | $i$ |
| $k$ | | $j\mathbf{i}$ | $j\mathbf{j}$ | $j\mathbf{k}$ | $-i$ | $-1$ |
| | | | | | | |

This is closer to the charge accommodation algebra, which is based on a quaternion-vector system. If we use a quaternion-vector system, as such, we obtain:

| | | $i\mathbf{i}$ | $i\mathbf{j}$ | $i\mathbf{k}$ | $j$ | $k$ |
|---|---|---|---|---|---|---|
| | | | | | | |
| $i\mathbf{i}$ | | $-1$ | $-i\mathbf{k}$ | $i\mathbf{j}$ | $k\mathbf{i}$ | $-j\mathbf{i}$ |
| $i\mathbf{j}$ | | $i\mathbf{k}$ | $-1$ | $-i\mathbf{i}$ | $k\mathbf{j}$ | $-j\mathbf{j}$ |
| $i\mathbf{k}$ | | $-i\mathbf{j}$ | $i\mathbf{i}$ | $-1$ | $k\mathbf{k}$ | $-j\mathbf{k}$ |
| $j$ | | $-k\mathbf{i}$ | $-k\mathbf{j}$ | $-k\mathbf{k}$ | $-1$ | $i$ |
| $k$ | | $j\mathbf{i}$ | $j\mathbf{j}$ | $j\mathbf{k}$ | $-i$ | $-1$ |
| | | | | | | |

This has the advantage of making all the diagonal terms identical, as we would require. Also, all the pure vector terms become complex, while the pure quaternion terms and the vector quaternion terms are not, which creates a greater degree of uniformity in the algebra. In this case, $i\mathbf{i}$, $i\mathbf{j}$, $i\mathbf{k}$ represents the strong charge in its three-colour form, $j$ the electric charge, and $k$ the weak charge. The terms $\pm\, i\mathbf{i}$, $\pm\, i\mathbf{j}$, $\pm\, i\mathbf{k}$ represent the colour non-singlet gluons; $\pm\, i$ the weak generators $W^{+}$ and $W^{-}$; and $\pm\, k\mathbf{i}$, $\pm\, k\mathbf{j}$, $\pm\, k\mathbf{k}$ and $\pm\, j\mathbf{i}$, $\pm\, j\mathbf{j}$, $\pm\, j\mathbf{k}$ the generators $X$ and $Y$ of the Grand Unified theory. It is significant now that the strong charge is represented by a quaternion $i$, which is 'privileged', by taking on the vector operators $\mathbf{i}$, $\mathbf{j}$, $\mathbf{k}$, but that the same quaternion, without the vector operators, represents the weak interacting generators, $W^{+}$ and $W^{-}$; while the colour non-singlet gluons are represented by complex pseudovectors, exactly as they are when represented as carried by the spin angular momentum.

Another possibility is to use a vector quaternion algebra which is close to the charge-accommodation form:

|  |  | *i*i | *i*j | *i*k | *j*j | *k*k |
|---|---|---|---|---|---|---|
|  |  |  |  |  |  |  |
| *i*i |  | 1 | – k | j | *ik*k | *ij*j |
| *i*j |  | k | 1 | – i | *k* | – *ij*i |
| *i*k |  | – j | i | 1 | – *ik*i | – *j* |
| *j*j |  | *ik*k | – *k* | – *ik*i | 1 | *ii*i |
| *k*k |  | *ij*j | – *ij*i | j | *ii*i | 1 |
|  |  |  |  |  |  |  |

or, alternatively, the equivalent double quaternion algebra:

|  |  | *i*i | *i*j | *i*k | *j*j | *k*k |
|---|---|---|---|---|---|---|
|  |  |  |  |  |  |  |
| *i*i |  | 1 | – k | j | *k*k | *jj* |
| *i*j |  | k | 1 | – i | – *k* | – *ji* |
| *i*k |  | – j | i | 1 | – *k*i | *j* |
| *j*j |  | *k*k | *k* | – *k*i | 1 | *ii* |
| *k*k |  | *jj* | – *ji* | j | *ii* | 1 |
|  |  |  |  |  |  |  |

The *s*-**p** connection provides a dynamical model of quarks using integral charges because, in all these case, the vector element in the mapping is preserved, irrespective of the sign. This is what makes it possible to write down a baryon wavefunction in terms of a 'rotating' **p**.

Within the context of the entire 32-part algebra, the group relationships may be shown in a $32 \times 32$ table, of which the first $12 \times 12$ products are:

| * | 1 | *i* | *ii* | *ij* | *i*k | *ik* | *j* | *j*i | *j*j | *j*k | *ii* | *k* |
|---|---|---|---|---|---|---|---|---|---|---|---|---|
| **1** | 1 | *i* | *ii* | *ij* | *i*k | *ik* | *j* | *j*i | *j*j | *j*k | *ii* | *k* |
| *i* | *i* | –1 | *iii* | *iij* | *ii*k | –*k* | *ij* | *ij*i | *ij*j | *ij*k | –*i* | *ik* |
| *ii* | *ii* | *iii* | –1 | –*i*k | *i*j | –*ij*i | *k*i | *k* | –*ik*k | –*ik*j | –*ii* | –*j*i |
| *ij* | *ij* | *ii*j | *i*k | –1 | –*ii* | –*ij*j | *k*j | *ik*k | *k* | *ik*i | –*ij* | –*j*j |
| *i*k | *i*k | *ii*k | –*ij* | *ii* | –1 | –*ij*k | *k*k | *ik*j | –*ik*i | *k* | –*i*k | –*j*k |
| *ik* | *ik* | –*k* | *ij*i | *ij*j | *ij*k | 1 | –*ii* | *ii* | *ij* | *i*k | –*j* | –*i* |
| *j* | *j* | *ij* | –*k*i | –*k*j | –*k*k | *ii* | –1 | –*i* | –*j* | –*k* | –*ik* | *i* |
| *j*i | *j*i | *ij*i | –*k* | –*ik*k | *ik*j | –*ii* | –*i* | –1 | –*i*k | *ij* | –*ik*i | *ii* |
| *j*j | *j*j | *ij*j | *ik*k | –*k* | –*ik*i | –*ij* | –*j* | *i*k | –1 | –*ii* | –*ik*j | *ij* |
| *j*k | *j*k | *ij*k | –*ik*j | *ik*i | –*k* | –*i*k | –*k* | –*ij* | *ii* | –1 | –*ik*k | *i*k |
| *ii* | *ii* | –*i* | –*ii* | –*ij* | –*i*k | *j* | *i*k | *ik*i | *ik*j | *ik*k | 1 | –*ij* |
| *k* | *k* | *ik* | *j*i | *j*j | *j*k | –*i* | –*i* | –*ii* | –*ij* | –*i*k | *ij* | –1 |

The remaining 20 rows and columns are the products of the four pentads (***k***i, ***k***j, ***k***k, *ij*, *i*); ( *ii*i, *ii*j, *ii*k, *i*k, **j**); ( *iji*, *ijj*, *ijk*, *ii*, **k**); ( *i***k**i, *i***k**j, *i***k**k, *ij*, **i**). The full group has 64 elements, constructed from the positive and negative versions of the 32 algebra units. For convenience, the multiplications of the negative elements are not shown, but the multiplications of two negative elements will generate the same products as the multiplications of two positive elements, while the multiplications of positive with negative elements, in either order, will generate the same products as two positive elements, if subsequently multiplied by –1. From the table it is clear that the 32-part algebra may be constructed from 1, *i*, and six independent Dirac pentads, which double to twelve with the signs reversed. Within each pentad, it is possible to observe the eight $SU(3)$ generators and the four generators for $SU(2)_L \times U(1)$, as outlined in the earlier part of this section.

In the table, the 'canonical' pentad (*ii*, *ij*, *i*k, *ik*, **j**), obtained by privileging the ***i*** quaternion operator, representing strong charge, is followed by two further pentads, obtained by cycling the quaternion operators and successively privileging ***j*** and ***k***, the operators representing the electromagnetic and weak charges. This demonstrates that the $SU(5) / U(5)$ arrangement of the 'quark' tables A-E represents three interlocking $SU(3)$ systems for the three interactions, of which only one may be privileged as the *physical* carrier of the vector aspect of angular momentum conservation. The existence of three further pentads, in which complex quaternion operators replace vectors, and vectors replace quaternions, is a consequence of the fact that our original creation of the Dirac algebra privileged the vectors by allowing them to retain their perfect symmetry at the expense of the quaternion operators. Mathematically, however, it would have been just as valid to privilege the quaternion operators at the expense of the vectors.

As a further example of pentad structure, it may be that the 5-fold symmetry of Penrose tiling can be related to the Dirac algebra, as it derives from combinations of two types of 4-sided figures: darts and kites. Each of these is an equilateral triangle, with one of the sides having a kink, either inwards or outwards, to produce the fourth vertex. In principle, this is based on 2 types of 3+1, and there may be a connection with quaternions and 4-vectors. The combination tends to produce figures with 5 or a multiple of 5 vertices on the outside. It is possible that there is a connection between this and the eight parts of space, time, mass and charge becoming the five of Dirac energy-momentum-mass. The 7-D mapping with 7 possible neighbourhoods for the tiling patterns may relate to the imaginary part of an octonion-type structure. It may also be relevant that Penrose tiling is fractal, as is the infinite series of pentad vacuum 'images' produced by a fermion.

## 24 Superspace and higher symmetries

We have seen that the whole set of groups relevant to the foundations of physics comes from the fundamental concept of duality, which is effectively the same thing as applying a $C_2$ symmetry to the idea of physical measurement, and which has direct expression in physical equations through the numerical factor 2. The symmetry group of the parameters develops from three $C_2$ symmetries, and extends to higher symmetries on application of the specific mathematical forms applicable to these dualities. Successive applications of the real / imaginary and discrete / continuous (or unidimensional / multidimensional) divisions between the parameters space, time, mass and charge leads to a fundamental group of order 64 (the Dirac algebra) applicable to the symmetry of an object. At the same time, the application of the conserved / nonconserved properties also produces the concept of rotational symmetry, which, applied to the multidimensionality which derives from the discrete / continuous division, introduces a related set of Lie algebras, specified by a finite number of generators rather than elements, including the subgroups of $G_2$, which relate to the Standard Model. It may be that we can continue to find meaning by doubling beyond this stage, and using the Freudenthal-Tits Magic Square. It is even possible that the doubling is open-ended, like that in Newton's third law, or the 'supersymmetric' creation in the vacuum of infinite numbers of fermionic and bosonic states. It may even be their actual physical representation.

Using the Magic Square, which includes the 5 unique order 8 abstract groups, or exceptional Lie groups, generated by the octonions, we may extend our analysis to groups such as $E_6$ and $E_8$, which have particularly interesting characteristics for physical theory. The $M_3^8$ or $3 \times 3$ Hermitian octonionic matrix representations of the complexified $E_6$, for instance (or $E_6 \times U(1)$, with $U(1)$ representing the phase term), which Gürsey et al have seen as a possible grand unification group uniting quarks and leptons,[38-40] have 27 degrees of freedom, which is comparable with the 27 possible particle tables of the form A-E. The group $E_8$, on the other hand, with 256 generators, 8 of which are 'timelike', which may be needed to complete the set of octonion-related symmetries, has connections with supersymmetry and string theories, and so it is particularly interesting that it is a natural product of the hierarchies created by space, time, mass and charge. The real $8 \times 8$ matrices of $E_8$ could possibly be isomorphic to the complex $4 \times 4$ matrices of the Dirac group.

In supersymmetry theories, the vacuum has zero energy if the symmetry is unbroken; in the present case the symmetry-breaking is due to the Dirac / Higgs mechanism, which privileges $+E$ states over $-E$. It is probably very significant that the superspace needed for supersymmetry postulates four antisymmetric coordinates as superpartners of space-time – these look very like the mass and three charges of the present theory. The eight coordinates together provide a superspace, which is like the nilpotent Dirac algebra or double quaternions. Perhaps significantly, $E$-**p**-$m$ (or $t$-**r**-$\tau$) and $w$-$s$-$e$ together provide a 10-dimensional possibility, the eleventh dimension of string theory being introduced to embed the 10-D required in an observable structure, like the 2-D needed to draw a line in space.

The present model does not need explanation in terms of string theory, but it may explain how such theories are generated, and perhaps produce new results at higher levels. The most popular group representation for the superstrings which are generalizations of supersymmetry is $E_8 \times E_8$. In this representation one $E_8$ forms the $SU(3) \times SU(2) \times U(1)$ symmetry, while the other $E_8$ set is assumed to be gravitational terms. We can speculate on the possibility of deriving all four forces from one set, with the other being a mirror set, thus connecting the real / virtual particle symmetry outlined in this paper to the formalism of string theory. Perhaps one set may represent space (3), time, mass, charge (3), combined as in $t$-**r**-$\tau$, and quantized via the charge input, and the other set Dirac momentum (3), Dirac energy, Dirac rest mass, and Dirac angular momentum (3). The second set contains all the conserved quantities related to mass, which may be connected to the association of the second $E_8$ with gravity – conservation of charge and angular momentum rotational / irrotational properties are certainly related, as we have seen, in the same way as those of space are related to momentum, and those of energy to time. Mass, also, is split between the two sets, the source being in one, and the 'field' being in the other, and this would seem to be in agreement with the spirit of supersymmetry.

## 25 Conclusion

The structure of fundamental physics appears to be based on a system of dualities seemingly designed to allow us to obtain 'something from nothing'. Nature, as a whole, remains uncharacterized and uncharacterizable when we define it in terms of the symmetrical fundamental parameter group space, time, mass and charge. The dualities within this set generate a full range of group structures, including both finite groups and the Lie groups defining their transformations, from the simplest finite group $C_2$ up to the exceptional Lie group $E_8$ involved in superstring theories. Incorporated within these group structures is a remarkably powerful version of the Dirac algebra, with a suggestive nilpotent structure, which, in turn, requires a broken symmetry between the four known interactions, and a mechanism for deriving fundamental particle structures and particle masses.